\documentclass{article}

\usepackage[nonatbib]{arxiv_final}

\usepackage[pdftex]{graphicx}

\usepackage{lineno}
\usepackage{hyperref}
\modulolinenumbers[5]

\usepackage[table]{xcolor}
\usepackage{arydshln}  
\usepackage{multirow}
\usepackage{mathtools, cuted}
\usepackage{amsmath}
\usepackage{amsfonts}
\usepackage{array}
\usepackage{url}
\usepackage{bm}
\usepackage{algorithmic}
\def\I{\mathrm{i}}

\makeatletter
\def\endthebibliography{%
	\def\@noitemerr{\@latex@warning{Empty `thebibliography' environment}}%
	\endlist
}
\makeatother

\begin{document}

\title{Characterising poroelastic materials in the ultrasonic range - A Bayesian approach}

\author{Matti~Niskanen$^{a,b,}$\thanks{Corresponding author.
		email: \emph{matti.niskanen@uef.fi}}\hspace{1ex},
	Olivier~Dazel$^{b}$,
	Jean-Philippe~Groby$^{b}$,
	Aroune~Duclos$^{b}$,
	Timo~L\"{a}hivaara$^{a}$}

\date{$^{a}$ Department of Applied Physics, University of Eastern Finland, Yliopistonranta 1, FIN-70211 Kuopio, Finland\\%
	$^{b}$ Laboratoire d'Acoustique de l'Universit\'{e} du Mans -- UMR CNRS 6613, Avenue Olivier Messiaen, \\F-72085 Le Mans Cedex, France\\[2ex]%
	\today}

\maketitle

\begin{abstract}
	Acoustic fields scattered by poroelastic materials contain key information about the materials' pore structure and elastic properties.
	Therefore, such materials are often characterised with inverse methods that use acoustic measurements.
	However, it has been shown that results from many existing inverse characterisation methods agree poorly.
	One reason is that inverse methods are typically sensitive to even small uncertainties in a measurement setup, but these uncertainties are difficult to model and hence often neglected.
	In this paper, we study characterising poroelastic materials in the Bayesian framework, where measurement uncertainties can be taken into account, and which allows us to quantify uncertainty in the results.
	Using the finite element method, we simulate measurements where ultrasonic waves are incident on a water-saturated poroelastic material in normal and oblique angles.
	We consider uncertainties in the incidence angle and level of measurement noise, and then explore the solution of the Bayesian inverse problem, the posterior density, with an adaptive parallel tempering Markov chain Monte Carlo algorithm.
	Results show that both the elastic and pore structure parameters can be feasibly estimated from ultrasonic measurements.
\end{abstract}

\section{Introduction}

Modelling the propagation of sound in poroelastic media is motivated by two complementary aims.
Firstly, it is needed to explain the acoustic absorption and transmission properties of structures.
Secondly, the sound waves interact with the material in a non-destructive manner and carry information about its micro-structure and elastic behaviour.
Knowledge of these properties is highly valuable in many areas such as geophysical exploration \cite{slatt2013stratigraphic}, design of materials for noise treatment \cite{allard2009propagation,cox2016acoustic}, industrial filtration applications \cite{espedal2007filtration}, and characterisation of bone tissues \cite{baroncelli2008quantitative}.

Inverse characterisation methods are based on conducting measurements that can be related to the unknowns of interest through a physical model.
It is possible to measure some material properties directly, but the use of such methods is limited and they are not considered in this paper.
The attractiveness of inverse methods for porous materials stems partly from the fact that they can rely on relatively simple acoustic measurements.
In the audible frequency range, such experiments are usually done with the impedance tube.
The tested material is cut into the shape of the tube and faces normally to the incident wave field.
For an overview of these methods we refer to recent review papers by Bonfiglio and Pompoli \cite{bonfiglio2013inversion} and Horoshenkov \cite{horoshenkov2017review} (see also \cite{zielinski2015normalized} and the references therein).
Another option is to do the measurements in an open field, typically in the ultrasonic frequency range, although audible frequencies can be used with large enough samples.
In the open field, the sound source can be pointed at the tested object either in normal or oblique incidence \cite{jocker2007minimization,groby2010analytical}.

Currently, numerous direct and inverse characterisation methods are in use, and the results they give do not agree very well.
This phenomenon has been noted in the case of rigid frame porous materials in two inter-laboratory tests by Horoshenkov \emph{et al.} in 2007 \cite{horoshenkov2007reproducibility} and Pompoli \emph{et al.} in 2017 \cite{pompoli2017reproducible}.
The reproducibility of the results was poor, especially with materials having a high flow resistivity.
Recently, Bonfiglio \emph{et al.} \cite{bonfiglio2018reproducible} reported a result where fourteen independent laboratories measured the elastic properties of specimens of the same porous materials.
Again the finding was that the reproducibility of the experimental methods is poor, and in the worst case the results differed by two orders of magnitude.
This poor reproducibility can be attributed to a lack of standardised measurement and calibration procedures, and to underlying uncertainties in measurement setup, which are difficult to model and therefore often neglected.
However, it has been shown that, in inverse problems, neglecting modelling errors typically yields meaningless estimates \cite{kaipio2013approximate,nicholson2018estimation}.

Most of the inverse methods found in literature for characterisation of porous media are deterministic and based on iteratively minimising a cost function to find a point estimate for the parameters.
A problem with the deterministic approach is that it does not provide a simple framework for quantifying uncertainty in the estimates.
Another issue is that the cost function often exhibits several minima, and finding the global minimum can be difficult \cite{aster2018parameter}.
Several papers have discussed ways to circumvent this problem, by the use of multiple cost functions formed from different frequency data \cite{ogam2011non}, normalising the estimated parameters before doing inversion \cite{zielinski2015normalized}, or using differential evolution algorithms \cite{atalla2005inverse}, among others.
A promising recently proposed approach is to minimise the cost function in a stepwise manner \cite{goransson2019parameter}.

In practice the models and measurement data always include uncertainties, and so the material parameters cannot be known exactly.
Therefore, we adopt the Bayesian framework (see for example \cite{kaipio2006statistical,calvetti2007introduction,gelman2013bayesian}), where the parameters are treated as random variables, all prior information on the parameters is coded into a prior probability density function (pdf), and the solution to the inverse problem is the posterior pdf.
The Bayesian approach can account for the structure and level of measurement noise as well as the effect of modelling errors.
The posterior density then allows the calculation of various point estimates but also credibility intervals, which represent the uncertainty in the estimates.
These can be calculated by sampling the posterior with Markov chain Monte Carlo (MCMC) methods.
Based on such analysis we can decide if the accuracy of the estimates is high enough for our application, or if more information is needed (for example in the form of additional measurements).

For the reasons outlined above, the Bayesian approach has gained popularity in the recent years.
To characterise porous media using the impedance tube, Bayesian methods have been considered both in the case where the solid part of the material (the frame) is modelled as rigid \cite{niskanen2017deterministic}, and where it is modelled as elastic \cite{chazot2012acoustical}.
The rigid frame case has also been considered in the free field ultrasonic regime \cite{roncen2018bayesian}.
In geoacoustics, Bayesian methods have been used to characterise the sediment structure of seabed, using Bayesian model selection and parameter inference \cite{dettmer2009model,dettmer2010trans,dosso2012parallel,bonomo2018comparison}.

In this work, we consider a slab of ceramic-like poroelastic material and study the feasibility of fully characterising it using ultrasonic measurements.
The frame of the considered material is relatively stiff, and a problem is that sound waves in air would not be able to deform it and we would not see any elastic behaviour.
Using an impedance tube to measure such materials is therefore not practical, and instead we consider free field ultrasound measurements in water.
To model the material and its acoustic response we use Biot's theory of poroelasticity \cite{biot1956theoryLow,biot1956theoryHigh,biot1962mechanics}.
The goal is then to estimate all the model parameters that are related to the poroelastic object.
We also emphasize the importance of taking sources of uncertainty in the measurements into account in producing reliable estimates of the material parameters.
For example, the measurement noise level is taken as an additional parameter to be estimated.
Since this paper deals mainly with the data analysis and inversion, we simulate the (noisy) measurement data numerically.
This allows us to compare the inverse solution to the known material parameters.

The degree of difficulty of sampling the posterior using MCMC is partly related to the dimensionality of the problem.
In the case of a rigid frame material, the posterior pdf is relatively low-dimensional and easy to sample.
This is because such materials can be modelled as fluids, and relatively few parameters are needed to describe them.
Sampling the posterior pdf is a lot more challenging when we use the full poroelastic model, primarily due to the added model complexity and number of parameters.
Unknowns related to the uncertainty in the measurements further increase the problem's dimensionality.
Therefore, in the present work, we use state of the art MCMC techniques, namely adaptive methods \cite{haario2001adaptive,andrieu2008tutorial} and parallel tempering \cite{earl2005parallel} to explore the posterior efficiently.
In addition, we develop a computationally fast and stable method to solve the poroelastic model.
The solution is based on the global matrix method \cite{knopoff1964matrix,lowe1995matrix}.
To avoid an inverse crime \cite{kaipio2006statistical}, we simulate the measurement data with the finite element method.

The paper is organised as follows.
In Section \ref{sec:Biot_numsolution} we describe the Biot model and our approach to solving it.
In Section \ref{sec:inverse_problem} we formulate the inverse problem and outline the used MCMC sampling methods.
Next, in Section \ref{sec:numerical_experiments} we describe the numerical experiment and apply the inversion method.
Results are shown in Section \ref{sec:results}, followed by a discussion in Section \ref{sec:discussion} before a conclusion in Section \ref{sec:conclusion}.

\section{The forward model}
\label{sec:Biot_numsolution}

\begin{figure}[t]
	\centering
	\includegraphics[width=0.5\linewidth]{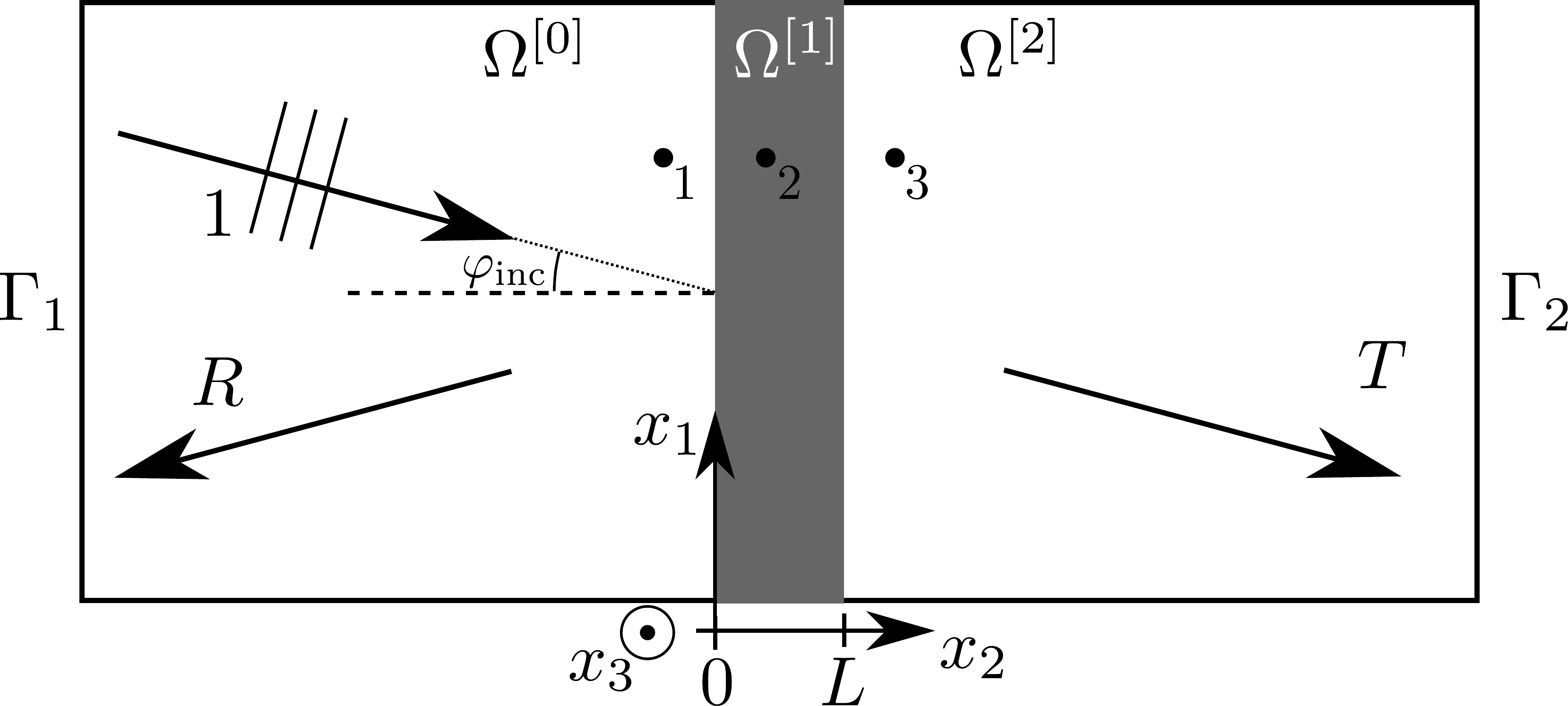}
	\caption{A schematic of the model geometry.}
	\label{fig:geometry_2D}
\end{figure}

We consider two fluid (water) domains $\Omega^{[0]}$ and $\Omega^{[2]}$, separated by a poroelastic slab $\Omega^{[1]}$ of thickness $L$ (see Fig.~\ref{fig:geometry_2D}).
The actual geometry extends to infinity in the $x_1$- and $x_3$-directions, and the boundaries surrounding the model are there only to truncate the computational domain for the finite element method calculations, described in Section \ref{sec:numerical_experiments}.
The forward problem consists of predicting the fields reflected and transmitted by the poroelastic layer in response to an incident plane wave impinging the structure with an angle $\varphi_{\mathrm{inc}}$ and initially propagating in $\Omega^{[0]}$.

The model is assumed to have circular symmetry around the $x_2$ axis, so that the coordinate system can be rotated to make the $(x_1,x_2)$ plane coincide with the sagittal plane defined by the incident wave vector.
Hence there is no propagation in the $x_3$-direction and the problem reduces to two spatial dimensions.

\subsection{Biot model of porous media}
\label{ssec:Biot_model}

Porous materials are made of a solid phase (called the frame) and of a fluid phase that is an interconnected network of pores inside the solid.
Here we assume that all of the pore volume is occupied by water.
When a flow of the fluid is able to cause the solid to deform, the material is called poro\emph{elastic}.
Poroelastic materials are most of the time modelled using the Biot theory.

The original Biot model \cite{biot1956theoryLow,biot1956theoryHigh} expresses the poroelastic medium in terms of the displacements of its homogenised solid phase $\mathbf{u}$ and fluid phase $\mathbf{U}$.
We adopt here an alternative formulation \cite{biot1962mechanics}, which describes the medium with the solid displacement and a fluid/solid relative displacement $\mathbf{w} = \phi(\mathbf{U} - \mathbf{u})$, where $\phi$ denotes the material's open porosity.
The alternative formulation is chosen because in our case it simplifies writing the interface conditions, which now do not include porosity, and because it is valid for inhomogeneous materials.
We consider the equations in the frequency domain, so an arbitrary time-dependent field $\bar{\chi}(\mathbf{x};t)$, $\mathbf{x} = (x_1,x_2)$, is related to its counterpart in the frequency domain with the Fourier transform $\chi(\mathbf{x};\omega) = \int_{-\infty}^{\infty}\bar{\chi}(\mathbf{x};t)e^{-\I \omega t} dt$.

In the alternative formulation Biot's motion equations read

\begin{align} \label{eq:Biot_motioneqn1}
& \omega^2\rho_f\mathbf{w} + \omega^2\rho\mathbf{u} = -\nabla\cdot\bm{\sigma},\\
& \omega^2\rho_f\mathbf{u} + \omega^2\tilde{\rho}_{\text{eq}}\mathbf{w} = \nabla\cdot p,\label{eq:Biot_motioneqn2}
\end{align}

\noindent where $\rho_f$ is the density of the fluid, $\rho = (1 - \phi)\rho_s + \phi\rho_f$ is the bulk density of the medium (with $\rho_s$ density of the solid), $\tilde{\rho}_{\mathrm{eq}}$ is the equivalent density of the porous material, $\bm{\sigma}$ is the total stress tensor, $p$ is the pressure, $\omega = 2\pi f$ is the angular frequency, and $f$ is the frequency.
For an isotropic porous material the stress-strain relations can be written as

\begin{align}
\bm{\sigma} & = 2N\bm{\epsilon} + (\lambda_c\nabla\cdot\mathbf{u} +\alpha_{\mathrm{B}} M\nabla\cdot\mathbf{w})\mathbf{I}, \label{eq:stress-strain1}\\
p & = -M(\nabla\cdot\mathbf{w} + \alpha_{\mathrm{B}}\nabla\cdot\mathbf{u}), \label{eq:stress-strain2}
\end{align}

\noindent where $N$ is the shear modulus, $\bm{\epsilon} = \frac{1}{2}\left(\nabla\mathbf{u} + (\nabla\mathbf{u})^T\right)$ is the strain tensor, and $\mathbf{I}$ is the identity tensor.
The parameter $\alpha_{\mathrm{B}}$ is the Biot-Willis coefficient \cite{biot1957elastic}, and $M$ and $\lambda_c = \lambda + \alpha_{\mathrm{B}}^2M$ are elastic parameters, where $\lambda$ is the first Lam\'{e}'s coefficient of the elastic frame.
These elastic parameters used by Biot can be stated as a function of $\phi, K_f, K_s$ (the bulk modulus of the elastic solid from which the frame is made of), and $K_b$ (the bulk modulus of the porous frame in a vacuum) with the relations

\begin{align}
\alpha_{\mathrm{B}} & = 1 - \frac{K_b}{K_s}, \\
M & = \left[\alpha_{\mathrm{B}} + \left(\frac{K_s}{K_f} - 1\right)\phi\right]^{-1}K_s, \\
\lambda_c & = \left[(1 + \phi)\alpha_{\mathrm{B}} + \left(\frac{K_b}{K_f} - 1\right)\phi\right]M - \frac{2}{3}N.
\end{align}

\noindent The set $\phi, K_f, K_s$, and $K_b$ can be measured directly, and we therefore consider them as fundamental properties of the material.

Viscous losses in the medium are accounted for in the equivalent density by the model of dynamic tortuosity introduced by Johnson \emph{et al.} \cite{johnson1987theory}.
Using this model we can write the equivalent density as

\begin{equation} \label{eq:JKD_rhoeq}
\tilde{\rho}_{\mathrm{eq}} = \frac{\rho_f}{\phi}\tilde{\alpha}(\omega),
\end{equation}

\noindent where the dynamic tortuosity is

\begin{equation}
\tilde{\alpha}(\omega) = \alpha_\infty + \frac{\I\nu}{\omega}\frac{\phi}{k_0}\sqrt{1 - \frac{\I\omega}{\nu}\left(\frac{2\alpha_\infty k_0}{\phi\Lambda}\right)^2} \:\:,
\end{equation}

\noindent where $\alpha_\infty$ is the geometric tortuosity, $k_0$ is the viscous static permeability, $\Lambda$ is the viscous characteristic length, and $\nu = \eta/\rho_f$ is the kinematic viscosity of the saturating fluid with $\eta$ the dynamic viscosity.

In addition to viscous losses introduced through the dynamic tortuosity model, we let the elastic parameters $K_b,K_s$, and $N$ be complex to allow for damping in the frame.
The parameters are assumed to be constant with respect to frequency, which corresponds to a slightly inelastic Biot model \cite{turgut1991investigation}.
This assumption is simpler than a fully poro-viscoelastic model, where the elasticity parameters are a function of frequency, but the constant parameter model has been shown to be slightly non-causal.
It can be shown, however, that when attenuation in the frame (corresponding to the ratio of imaginary to real part) is small, the elastic parameters can be approximated as constants \cite{bourbie1987acoustics}.

We seek to express the Biot motion equations \eqref{eq:Biot_motioneqn1} and \eqref{eq:Biot_motioneqn2} as a system of first order ordinary differential equations (ODEs) that can then be solved in several ways.
A common way is to use the transfer matrix method (TMM) which, while straightforward to implement, suffers from known numerical instabilities \cite{lowe1995matrix} at high frequencies and certain parameter combinations.
These issues make the TMM unsuitable for our use because the inversion algorithm solves the model over a wide range of parameters, and the numerical problems are often present.
Instead, we use the global matrix method (GMM) \cite{knopoff1964matrix}, which is stable even for high frequencies (as shown for example in \cite{lowe1995matrix,chin1984matrix,schmidt1986efficient}), and has also proved stable in our tests.
In the GMM, each backward and forward propagating wave inside the poroelastic material is explicitly written out as a function of their amplitudes, and their origin is individually chosen to be on the boundary from which they originate.

\subsection{State vector formalism}

We perform a Fourier transform of the field along $x_1$, which due to the plane wave nature of the excitation can be written as $\chi(\mathbf{x};\omega) = \hat{\chi}(k_1,x_2;\omega)e^{\I k_1x_1}$, where $k_1 = -k^{[0]}\sin (\varphi_{\mathrm{inc}})$, and $k^{[0]} = \omega/c_f$.
For the remainder of this paper we omit the $\omega$ and $k_1$ dependence in the expressions for the fields and write $\hat{\chi}(k_1,x_2;\omega) \equiv \hat{\chi}(x_2)$.

To write out the motion equations as an ODE system we introduce a state vector $\hat{\mathbf{s}}(x_2)$.
Components of the state vector can be selected arbitrarily as long as they completely describe the state of the system at $x_2$.
A convenient choice is to use the normal components of the stress tensor, $\hat{\sigma}_{12}^{[1]}$ and $\hat{\sigma}_{22}^{[1]}$, the pressure $\hat{p}^{[1]}$, the solid displacements $\hat{u}_1^{[1]}$ and $\hat{u}_2^{[1]}$, and the normal component of the total velocity $\hat{v}_2^{[1]} = -\I\omega(\hat{w}_2^{[1]} + \hat{u}_2^{[1]})$.
This choice simplifies the coupling conditions between interfaces in the current formulation of the Biot equations.
We therefore have $\hat{\mathbf{s}}(x_2) = [\hat{\sigma}_{12}^{[1]},\: \hat{\sigma}_{22}^{[1]},\: \hat{p}^{[1]},\: \hat{u}_1^{[1]},\: \hat{u}_2^{[1]},\: \hat{v}_2^{[1]}]$, where the $x_2$-dependence of the parameters has been dropped for clarity.

After performing the spatial Fourier transform, the motion equations \eqref{eq:Biot_motioneqn1}-\eqref{eq:Biot_motioneqn2} and the stress-strain relations \eqref{eq:stress-strain1}-\eqref{eq:stress-strain2} can be cast in the form

\begin{equation} \label{eq:ODEsystem}
\frac{\mathrm{d} \hat{\mathbf{s}}(x_2)}{\mathrm{d} x_2} + \mathbf{A}\hat{\mathbf{s}}(x_2) = 0.
\end{equation}

\noindent The matrix $\mathbf{A}$ is called the state matrix, and it is a function of the material parameters, components of the incident wave, and frequency.
The state matrix for this specific state vector was derived by Gautier \emph{et al.} \cite{gautier2011propagation} and is provided in \ref{app:statematrix}.

Four boundary conditions are needed to couple waves that propagate between a fluid and a poroelastic material \cite{allard2009propagation}.
Let $\bar{x}$ be a location of the interface separating the two mediums and let the superscript $0$ represent here either of the fluid domains $\Omega^{[0]}$ or $\Omega^{[2]}$.
The coupling conditions are continuity of pressure, $\hat{p}^{[1]}(\bar{x}) = \hat{p}^{[0]}(\bar{x})$, continuity of the normal component of velocity, $\hat{v}_2^{[1]}(\bar{x}) = \hat{v}_2^{[0]}(\bar{x}) = -\frac{\I}{\omega\rho_f}\frac{\mathrm{d} \hat{p}^{[0]}(x_2)}{\mathrm{d} x_2}\Bigr|_{\substack{x_2 = \bar{x}}}$, and continuity of the normal components of the stress tensor, $\hat{\sigma}_{12}^{[1]}(\bar{x}) = 0$ and $\hat{\sigma}_{22}^{[1]}(\bar{x}) = -\hat{p}^{[0]}(\bar{x})$.

In the fluid domains we can represent the incident, reflected, and transmitted waves as

\begin{subequations}
	\begin{align}
	\hat{p}^{[0]}(x_2) & = e^{-\I k_2x_2} + Re^{\I k_2x_2}, \\
	\hat{p}^{[2]}(x_2) & = Te^{-\I k_2(x_2 - L)},
	\end{align}
\end{subequations}

\noindent where $k_2 = k^{[0]}\cos (\varphi_{\mathrm{inc}})$.
Substituting these into the coupling relations we get on the first interface

\begin{subequations}
	\begin{align}
	\hat{p}^{[1]}(0) & = 1 + R, \\
	\hat{v}_2^{[1]}(0) & = -\frac{1}{Z_f}\cos(\varphi_{\mathrm{inc}}) + \frac{R}{Z_f}\cos(\varphi_{\mathrm{inc}}), \\
	\hat{\sigma}_{12}^{[1]}(0) & = 0, \\
	\hat{\sigma}_{22}^{[1]}(0) & = -1 - R,
	\end{align}
\end{subequations}

\noindent where $Z_f = \rho_fc_f$ is the characteristic impedance of the fluid. A similar system is written on the second interface.

\subsection{Global matrix solution of the Biot equations}

In an isotropic poroelastic layer, there are three forward propagating waves and three backward propagating waves.
The idea in the GMM is to express the physical wave fields as a function of the wave amplitudes.
Then we can choose to represent the origin of the wave at the interface it originates from and hence avoid the numerical issues that occur in the TMM.
After applying the coupling conditions between layers, we can solve the transmission and reflection coefficients.

The solution proceeds as follows \cite{dazel2013stable}.
We first calculate the eigendecomposition of $\mathbf{A}$:

\begin{equation} \label{eq:alpha_eigDecomposition}
\mathbf{A} = \mathbf{\Phi}\mathbf{\Gamma}\mathbf{\Phi}^{-1},
\end{equation}

\noindent where the column $\mathbf{\Phi}_j$ of the $(6\times6)$ eigenvector matrix $\mathbf{\Phi}$ corresponds to the polarisation of the $j$-th wave, and the eigenvalue $\mathbf{\Gamma}_{jj}$ on the diagonal corresponds to $\I k^{[1]}_j$, $j = 1,\dots,6$ \cite{dazel2013stable,de2009materials}.
The eigenvalues are ordered in pairs of a forward and a backward propagating wave.
A classical solution to \eqref{eq:ODEsystem} is $\hat{\mathbf{s}}(x_2) = \exp\{-(x_2 - \mathbf{x}_*)\mathbf{A}\}\hat{\mathbf{s}}(\mathbf{x}_*)$, which, using decomposition \eqref{eq:alpha_eigDecomposition}, can be written as

\begin{equation}
\hat{\mathbf{s}}(x_2) = \mathbf{\Phi}\mathbf{\Lambda}(x_2)\mathbf{\Phi}^{-1}\hat{\mathbf{s}}(\mathbf{x}_*),
\end{equation}

\noindent where $\mathbf{\Lambda}(x_2) = \exp\{-(x_2 - \mathbf{x}_*)\mathbf{\Gamma}\}$ is the propagation factor along $x_2$.
The reference location $\mathbf{x}_*$ is set at the first interface $(\mathbf{x}_* = 0)$ in the TMM, whereas the GMM allows us to change the basis of the waves so that their origin is on the interface they originate from.
Let us define the new origin as $(j=1,\dots,6)$

\begin{equation}
x_{*,j} = \begin{cases}
0, \quad\: j \:\: \mathrm{odd} \\
L, \quad j \:\: \mathrm{even}.
\end{cases}
\end{equation}

\noindent The odd index refers to a forward, and even to a backward propagating wave.
Next, let $\mathbf{q} = \mathbf{\Phi}^{-1}\hat{\mathbf{s}}(\mathbf{x}_*)$, so we can express the state vector as

\begin{equation} \label{eq:eigenvalue_amplitudes}
\begin{split}
\hat{\mathbf{s}}(x_2) & = \mathbf{\Phi}\mathbf{\Lambda}(x_2)\mathbf{q} \\
& =: \mathbf{M}(x_2)\mathbf{q}.
\end{split}
\end{equation}

\noindent Note that the $(6\times1)$ vector $\mathbf{q}$ does not depend on $x_2$.
This allows us to combine two systems of the form \eqref{eq:eigenvalue_amplitudes}, with a common $\mathbf{q}$ but the state vector evaluated at different interfaces, to create the global matrix.

Before stacking the equations together, we reduce redundant degrees of freedom to speed up the calculations.
Each row in the $(6\times6)$ matrix $\mathbf{M}(x_2)$ is related to the corresponding state vector variable.
Therefore, to relate the four coupling conditions on both sides of the material, we do not need the fourth and fifth rows corresponding to $u_1$ and $u_2$.
Let us denote the $(4\times6)$ matrix with these rows removed by $\mathbf{M}^{-}(x_2)$, and a reduced state vector (without $u_1$ and $u_2$) by $\hat{\mathbf{s}}^{-}(x_2)$.

Now we can write out the state vectors and corresponding matrices on both interfaces, and concatenate the system as

\begin{equation}
\begin{bmatrix}
\hat{\mathbf{s}}^{-}(0) \\
\hat{\mathbf{s}}^{-}(L) \\
\end{bmatrix}
=
\begin{bmatrix}
\mathbf{M}^{-}(0) \\
\mathbf{M}^{-}(L)
\end{bmatrix}
\mathbf{q}.
\end{equation}

\noindent Substituting in the coupling conditions we get the following linear system with 8 equations and 8 unknowns

\begin{equation} \label{eq:RTsystem}
\begin{bmatrix}
0\\
-1\\
1\\
-\frac{\cos(\varphi_{\mathrm{inc}})}{Z_f}\\
0\\
0\\
0\\
0
\end{bmatrix}
=
\left[
\begin{array}{c;{2pt/3pt}c;{2pt/3pt}c}
0 & & 0 \\
1 & \mathbf{M}^{-}(0) & 0 \\
-1 & & 0 \\
-\frac{\cos(\varphi_{\mathrm{inc}})}{Z_f} & & 0 \\ \hdashline[2pt/3pt]
0 & & 0 \\
0 & \mathbf{M}^{-}(L) & 1 \\
0 & & -1 \\
0 & & \frac{\cos(\varphi_{\mathrm{inc}})}{Z_f}
\end{array}
\right]
\begin{bmatrix}
R \\
\mathbf{q} \\
T
\end{bmatrix},
\end{equation}

\noindent from which the reflection and transmission coefficients are easy to solve.

The system \eqref{eq:RTsystem} must be solved for each frequency at a time, and one loop over the frequencies constitutes as one ''forward solution''.
To loop over the frequency range faster, we can stack the 8x8 matrices of each frequency on the diagonal of a bigger matrix.
This block diagonal matrix is typically very sparse and fast algorithms exist to solve the system for all the frequencies at once.
Computing one forward solution with 100 frequencies takes between 5 and 10 milliseconds using MATLAB\textsuperscript{\textregistered} R2018a on a laptop with an Intel i7-4710MQ processor.

\section{The inverse problem}
\label{sec:inverse_problem}

In the Bayesian framework all unknown parameters are modelled as random variables, and the inverse problem is seen as a problem of statistical inference \cite{kaipio2006statistical,calvetti2007introduction}.
The solution of a Bayesian inverse problem is the posterior pdf, which is constructed based on measured data, a model relating the measurements to the unknowns, and any prior information that might be available.
The posterior pdf represents all the information we have on the unknowns.

Let us denote the observable quantities by $\mathbf{y}$, the unknowns by $\bm{\theta}$, and measurement error (noise) by $\mathbf{e}$.
To keep the notations simple, we will denote random variables and their fixed realisations with the same symbol.
Our data vector $\mathbf{y}\in\mathbb{C}^{2n_\omega}$ consists of the reflection and transmission coefficients solved over $n_\omega$ frequencies.
For the exact definition of $\mathbf{y}$, see Sec.~\ref{ssec:FEM_data}.
Further, let us denote the number of unknowns by $n_\theta$ so that $\bm{\theta}\in\mathbb{R}^{n_\theta}$.
Assuming additive noise, we can model the measurement as

\begin{equation} \label{eq:observation model}
\mathbf{y} = h(\bm{\theta}) + \mathbf{e} \:\:,
\end{equation}

\noindent where $h: \mathbb{R}^{n_\theta} \rightarrow \mathbb{C}^{2n_\omega}$ is the forward model (the GMM solution over multiple frequencies) that maps the unknown parameters to measurable data, and $\mathbf{e}\in\mathbb{C}^{2n_\omega}$ denotes measurement noise.

According to Bayes' formula, the posterior density $\pi(\bm{\theta}|\mathbf{y})$ is proportional (up to a normalising constant) to the product of a likelihood $\pi(\mathbf{y}|\bm{\theta})$ and prior $\pi(\bm{\theta})$ pdfs

\begin{equation} \label{eq:Bayes}
\pi(\bm{\theta}|\mathbf{y}) = \frac{\pi(\mathbf{y}|\bm{\theta})\pi(\bm{\theta})}{\pi(\mathbf{y})} \propto \pi(\mathbf{y}|\bm{\theta})\pi(\bm{\theta}) \:\:.
\end{equation}

The prior density is constructed to give a high probability to parameter values we expect to find, and a low probability for those we think are unlikely.
Such information can be based on for example other experiments or an expert's beliefs.
We can also use the prior to rule out values inconsistent with physical reality, by setting the probability of such parameter values to zero.

When the measurement noise and the unknowns are mutually independent, the likelihood density can be written as $\pi(\mathbf{y}|\bm{\theta}) = \pi_e(\mathbf{y} - h(\bm{\theta}))$, where $\pi_e$ denotes the pdf of $\mathbf{e}$.
We assume that the measurement noise is normally distributed with an expected value of zero and a covariance $\mathbf{\Gamma}_e$.
Then, the posterior density \eqref{eq:Bayes} becomes

\begin{equation} \label{eq:posterior}
\pi(\bm{\theta}|\mathbf{y}) \propto \exp\left\{-(\mathbf{y} - h(\bm{\theta}))^T\mathbf{\Gamma}_e^{-1}(\mathbf{y} - h(\bm{\theta}))\right\}\pi(\bm{\theta}).
\end{equation}

\noindent We further assume that the covariance $\mathbf{\Gamma_e}$ is diagonal, i.e. the measurements in $\mathbf{y}$ are statistically independent.
If we do not know the noise covariance $\mathbf{\Gamma}_e$, it is possible in the Bayesian framework to consider $\mathbf{\Gamma}_e$ as an additional unknown (or unknowns) to be estimated.
In such a case the posterior takes the form $\pi(\bm{\theta},\mathbf{\Gamma}_e|\mathbf{y})\propto\pi(\mathbf{y}|\bm{\theta},\mathbf{\Gamma}_e)\pi(\bm{\theta})\pi(\mathbf{\Gamma}_e)$, where $\pi(\mathbf{\Gamma}_e)$ represents a prior for the noise covariance.
Now, the normalising factor of the noise distribution is no longer constant and we have to include it in the expression of the likelihood.
By denoting $\mathbf{L}_e^T\mathbf{L}_e^{} = \mathbf{\Gamma}_e^{-1}$,
the posterior is written as

\begin{equation} \label{eq:posterior_withnoise}
\pi(\bm{\theta},\mathbf{\Gamma}_e|\mathbf{y}) \propto \exp\left\{-\left\Vert \mathbf{L}_e(\mathbf{y} - h(\bm{\theta}))\right\Vert^2 -\log(\det(\mathbf{\Gamma}_e))\right\}\cdot\pi(\bm{\theta})\pi(\mathbf{\Gamma}_e).
\end{equation}

\subsection{Calculating the parameter estimates}

In order to explore the posterior density to calculate parameter and uncertainty estimates, we need a way to draw samples from it.
For this purpose we employ MCMC methods \cite{brooks2011handbook}, which are a widely used family of algorithms that generate correlated samples from the posterior.
MCMC methods are especially suitable for computing the conditional mean (CM) estimate

\begin{equation} \label{eq:CMestimate}
\hat{\bm{\theta}}_{\mathrm{CM}} = \int_{\mathbb{R}^{n_\theta}}\bm{\theta}\pi(\bm{\theta}|\mathbf{y})d\bm{\theta} \approx \frac{1}{n}\sum_{i=1}^{n}\bm{\theta}^{(i)},
\end{equation}

\noindent where $\{\bm{\theta}^{(i)}\}$ is an ergodic Markov chain produced by the MCMC sampler, and $n$ is the sample size.
The approximation in \eqref{eq:CMestimate} becomes exact in the limit $n\rightarrow\infty$.
From the posterior samples we can also compute uncertainty estimates, such as credible intervals.
A 95 \% credible interval $I_k(95) = [a_I,b_I]\subset\mathbb{R}$ for the parameter $\theta_k$ is defined as

\begin{equation}
\int_{a_I}^{b_I}\pi(\theta_k|\mathbf{y})d\theta_k = 0.95,
\end{equation}

\noindent where $\pi(\theta_k|y) = \int_{\mathbb{R}^{n_\theta-1}}\pi(\theta_1,\dots,\theta_{n_\theta}|\mathbf{y})d\theta_1\cdots\theta_{k-1}\theta_{k+1}\cdots\theta_{n_\theta}$ is the marginal density of the $k$-th component $\theta_k$ of $\bm{\theta}$.
In this paper we consider the narrowest interval $I_k(95)$.
The Bayesian credible intervals can be directly interpreted as probability statements about the likely values of the parameters given the data \cite{gelman2013bayesian}.

A downside to MCMC is that it is computationally very demanding since generating posterior samples relies on repeatedly solving the forward model.
In high-dimensional cases, and with computationally demanding forward models, running MCMC quickly becomes infeasible and often in such a case the only option is to solve the maximum a posteriori (MAP) estimate, which is an optimization problem.
To make sampling and the calculation of the CM estimate viable in our problem, we make use of adaptive MCMC methods and parallel tempering.
In general, the reason for using adaptive methods and parallel tempering is to improve the sampling efficiency by reducing autocorrelation times and facilitating easier jumping between posterior modes.
We describe the algorithm next, and then later use it for the parameter estimation problem.

\subsection{An adaptive parallel tempered MCMC algorithm}

The Metropolis-Hastings (M-H) algorithm \cite{metropolis1953equation,hastings1970monte} is the standard tool for drawing samples from a target probability distribution $\pi(\cdot)$.
In the random walk Metropolis algorithm, given the current location $\bm{\theta}^{(i)}$ of the chain, a candidate $\bm{\theta}^*$ for the next step is drawn from a proposal probability distribution that is symmetric and centered around $\bm{\theta}^{(i)}$.
Then the proposed location is either accepted or rejected with an acceptance probability $P(\bm{\theta}^{(i)},\bm{\theta}^*) = \min\{1,\pi(\bm{\theta}^*)/\pi(\bm{\theta}^{(i)})\}$.
For reasons of numerical stability, it is preferable to work with the logarithm of the target distribution so that the acceptance ratio becomes $\log\pi(\bm{\theta}^*)-\log\pi(\bm{\theta}^{(i)})$.

A usual choice for the proposal density is a multivariate Gaussian.
The candidate for the next move is thus drawn from $\bm{\theta}^* \sim \mathcal{N}(\bm{\theta}^{(i)},\mathbf{\Sigma})$, where $\mathbf{\Sigma}$ is the proposal covariance.
In an $n_\theta$ dimensional problem $\mathbf{\Sigma}$ has up to $n_\theta(n_\theta+1)/2$ tunable variables.
Correct tuning of the proposal covariance is crucial for the efficiency of the M-H algorithm, but tuning the proposal for a low-dimensional problem by hand is difficult and for large $n_\theta$ it is practically impossible.

An alternative to hand-tuning the covariance is to learn it on the go, using adaptive algorithms such as the Adaptive Metropolis (AM) proposed by Haario \emph{et al.} \cite{haario1999adaptive,haario2001adaptive}.
The idea of AM is that, during sampling, the covariance $\mathbf{\Sigma}$ is continuously updated based on the MCMC samples accumulated so far.
In this way the proposal size and orientation is automatically adapted to the shape of the target density, which is beneficial especially when some parameters are correlated.
At iteration $i$, we denote the covariance by $\mathbf{\Sigma}_i$, and the proposal density is of the form $\mathcal{N}(\bm{\theta}^{(i)},s_d\mathbf{\Sigma}_i)$, where $\mathbf{\Sigma}_i = \mathrm{cov}(\bm{\theta}^{(1)},...,\bm{\theta}^{(i)})$ and $s_d$ is a scaling parameter.
As a rule of thumb one can set $s_d = 2.4^2/n_\theta$ \cite{gelman1996efficient}.
With non-Gaussian target densities and at the beginning of the MCMC run when the estimate of the posterior covariance is uncertain, the scaling parameter $s_d$ can be too large which results in a low acceptance ratio and efficiency.
To improve this, Andrieu and Thoms \cite{andrieu2008tutorial} proposed to also adapt $s_d$ to achieve the desired acceptance rate.
We include this adaptation with a target acceptance ratio of 0.2.

In general, the M-H algorithm works well when $\pi(\cdot)$ is close to a unimodal distribution, but when the target distribution consists of multiple separated modes it tends to get stuck in one of them and fail to sample the other modes.
We observed this behaviour in the current parameter estimation problem, where the random walk chains adapted to a local maximum and in practice never found the main mode (whose location is known in simulated problems) unless the chain was started in it.

Tempering, i.e. considering a target density $\pi(\cdot)^{1/T}$, where $T\geq 1$ is called the temperature, flattens and widens the modes thus enabling a Markov chain to move around the density more freely.
In parallel tempering (PT) \cite{swendsen1986replica,geyer1991markov,earl2005parallel} we use $m$ Markov chains to sample in parallel $m$ tempered densities with temperatures $1 = T_1 < T_2 < \cdots < T_m$, and couple the chains together to exchange information on their location in parameter space.
In this way, the (possibly very isolated) high-probability locations found by the hot chains eventually propagate all the way to the chain at $T_1$, and as a result the cold chain is able to effectively explore the whole parameter space.
Note that the additional chains are introduced only to improve sampling efficiency of the chain at $T_1$, and at the end of the run only the chain at $T_1$ is retained.

In this work we consider a particular form of tempering where the likelihood is tempered but the prior is not:

\begin{equation}
\pi_{T}(\bm{\theta}|\mathbf{y}) \propto \pi(\mathbf{y}|\bm{\theta})^{\frac{1}{T}}\pi(\bm{\theta}),
\end{equation}

\noindent and set $T_m = \infty$.
At the highest temperature the target density corresponds to the prior only, which by our construction is unimodal and thus easy to sample.
Restricting the hottest chain to parameter values supported by the prior is a natural limit for the volume of parameter space we want to search in, because values outside the prior support we have, by definition, deemed impossible.

Coupling the chains together is achieved by proposing at pre-defined intervals to swap the locations of pairs of chains in the parameter space.
The swap is accepted with a probability that preserves the joint distribution of the chains.
This probability can be derived directly from the Metropolis update criterion, and in the case where we only temper the likelihood, it can be stated as \cite{vousden2015dynamic}

\begin{equation}
P_{i,j} = \min\left\{1, \left(\frac{\pi(\mathbf{y}|\bm{\theta}_i)}{\pi(\mathbf{y}|\bm{\theta}_j)}\right)^{1/T_j - 1/T_i}\right\},
\end{equation}

\noindent where $\bm{\theta}_i$ is the location in the parameter space of the $i$th chain and $T_i$ is the chain's temperature.
Usually the pairs that are adjacent in temperature are selected, because the swap acceptance probability decays quickly with a growing temperature difference.

Simulating $m$ chains obviously increases the computational demand $m$-fold.
However, this can be offset by a more than $m$-fold increase in sampling efficiency in strongly nonlinear and multimodal problems.
The efficiency of the PT sampler is greatly affected by the between chain acceptance probabilities, in a way similar to the within-chain acceptance probability issues of the M-H algorithm.
The between chain acceptance ratio can be tuned by adjusting the temperatures.
The closer any two chains are in temperature, the higher the probability of swapping them is, and vice versa.
The optimal acceptance rate for temperature swaps turns out to be the same as the optimal M-H acceptance rate, around 0.23 \cite{kone2005selection}.
To achieve this we start with $m$ geometrically spread out temperatures, and fix $T_1 = 1$ and $T_m = \infty$.
Temperatures $T_2,...,T_{m-1}$ are continuously adapted as the simulation goes on, following the procedure in Vousden \emph{et al.} \cite{vousden2015dynamic}.
This method changes the temperatures in such a way that an equal acceptance rate is achieved between each temperature pair.

Finally, we note that the number of temperatures $m$ could be automatically adjusted during an MCMC run, but to simplify the algorithm we have found it is enough to set $m$ manually before the start of the run.
If all acceptance rates between chains are too low ($< 0.15$), we restart the run with more temperatures.
In the case that the overall temperature swap rates are too high ($> 0.4$) we reduce the number of temperatures to avoid redundant sampling.

The following summarises the MCMC algorithm we have developed for use in this problem.
For clarity the considerations regarding burn-in and frequency of the adaptations have been omitted.
Swapping of the states is proposed at every iteration.

\begin{algorithmic}
	\STATE{\bfseries Adaptive PT algorithm}
	\STATE For $T=1,\dots,m$, initialise $\bm{\theta}^{(i,T)}$, $\mathbf{\Sigma}_i^{(T)}$, and $s_d^{(T)}$. Set $i=1$.
	\WHILE{$not$ $converged$}
	\FOR{$T = 1,\dots,m$}
	\STATE - Generate a candidate $\bm{\theta}^{(*,T)}$ and accept it with probability based on the Metropolis ratio.
	\STATE - Adapt $s_d^{(T)}$ based on the acceptance probability, and $\mathbf{\Sigma}_i^{(T)}$ based on the new sample.
	\ENDFOR
	\FOR{$T = 1,\dots,m-1$}
	\STATE - Swap the states between chains in adjacent temperatures with probability based on the Metropolis ratio.
	\ENDFOR
	\STATE - Adapt the temperatures $T_2,\dots,T_{m-1}$ based on the swap probabilities.
	\STATE - Check convergence
	\STATE - Set $i = i + 1$
	\ENDWHILE
\end{algorithmic}

\subsection{Convergence and Monte Carlo error}

Diagnosing convergence of a Markov chain is a subtle issue.
To avoid terminating the MCMC run prematurely or running it for an unnecessarily long amount of time, we implement an objective stopping criterion.
The chosen criterion is based on comparing Monte Carlo error in the CM estimate \eqref{eq:CMestimate} at iteration $n$, to the overall variability of the parameters.
However, we must keep in mind that no criterion can guarantee convergence and it is good practise to go through the usual checks (autocorrelation times, visual inspection of stationarity of the chains) every time as well.

Based on the central limit theorem (CLT) \cite{brooks2011handbook}, we can estimate the Monte Carlo standard error of $\hat{\bm{\theta}}_{\mathrm{CM}}$ with $\hat{\bm{\sigma}}_{\theta_{CM}}/\sqrt{n}$.
The estimate $\hat{\bm{\sigma}}_{\theta_{CM}}^2$ of the asymptotic variance in the CLT can be calculated via the consistent batch means method (see Flegal \emph{et al.} \cite{flegal2008markov}).
Half width of the $(1 - \alpha)\cdot 100$ \% confidence interval for the CM estimate is given by

\begin{equation}
\mathrm{\bf{CI}}_{1-\alpha} = t_{a_n-1}\left(1-\frac{\alpha}{2}\right)\frac{\hat{\bm{\sigma}}_{\theta_{CM}}}{\sqrt{n}} \:,
\end{equation}

\noindent where $t_{a_n-1}$ denotes the suitable quantile from Student's $t$-distribution with $a_n-1$ degrees of freedom, and $a_n$ is the number of batches used for the estimate $\hat{\bm{\sigma}}_{\theta_{CM}}^2$.

Our stopping criteria is then the following.
After a burn-in period, we first run the chain for long enough so that $\hat{\bm{\sigma}}_{\theta_{CM}}$ can be reliably estimated, and then at every $k$ iterations ($k = 1000$ for example), we compute $\mathrm{CI}_{95}$ for the CM estimate and the standard deviation $\bm{\sigma}_{\theta}$ of the samples accumulated so far.
When the uncertainty in the CM estimate gets below 10 \% of the parameter standard deviation, $\mathrm{\bf{CI}}_{95} < \bm{\sigma}_{\theta}/10$, for each parameter, the simulation is stopped.

\section{Numerical experiments}
\label{sec:numerical_experiments}

To investigate the performance of the inversion algorithm we conduct numerical experiments with data from a finite element simulation.
We simulate a material whose properties resemble those of commercially available filter materials made by sintering glass powder.
Properties of such a material can be found for example in Jocker \emph{et al.} \cite{jocker2009ultrasonic} (material S3) and they are shown in Table \ref{tab:inversion_results}.
We set the imaginary parts of the elastic coefficients to two percent of the real parts.
Properties of the fluid saturating the porous frame are: density $\rho_f = 1000$ kg$\cdot$m$^{-3}$, dynamic viscosity $\eta = 1.14\cdot 10^{-3}$ Pa$\cdot$s, speed of sound $c_f = 1500$ m$\cdot$s$^{-1}$, and bulk modulus $K_f = 2.19\cdot 10^9$ Pa.
For the considered material the viscous-inertial regime transition frequency is $\omega_c = \eta\phi/(\rho_f\alpha_\infty k_0) \approx 45 $ kHz.
To be well within the inertial regime we choose to work in the range 100 - 400 kHz.

\subsection{The finite element model and data}
\label{ssec:FEM_data}

The finite element model is constructed using the Pressure Acoustics and Poroelastic Waves interfaces in COMSOL Multiphysics\textsuperscript{\textregistered} software v. 5.3.
The model configuration is similar to Fig. \ref{fig:geometry_2D}.
Boundary $\Gamma_1$ is given a plane wave radiation condition, and the opposing boundary $\Gamma_2$ an absorbing impedance condition.
The impedance boundary is known to work perfectly only at normal incidence, and reflect some of the energy back when hit by a wave propagating at an oblique angle.
We use the impedance boundary condition as a way to introduce model discrepancy that is not accounted for in the inversion.
To approximate plane wave propagation, we take a 15 mm wide (in the $x_1$-direction) slice of the material as a ''unit cell'', and extend it to infinity by giving the upper and lower sides of the geometry a periodic Floquet condition.
The interior boundaries from left to right have the acoustic-poroelastic and poroelastic-acoustic condition, respectively.
The mesh consists of 2,760 quadrilateral elements.
All elements have the same size 0.5 mm, which corresponds to 6 quadratic elements per wavelength in water at 500 kHz.
We consider the frequency range 100 - 400 kHz with a 5 kHz resolution, so in total there are 61 frequencies in the data.
We simulate measurements where the incident wave is at an angle between $0^{\circ}$ and $45^{\circ}$.

The parametrisation of the Biot model in COMSOL is otherwise similar to our implementation, except for the way the equivalent density \eqref{eq:JKD_rhoeq} is written.
We choose to use different viscous loss models to further avoid using a too similar model in the inversion.
COMSOL implements the original loss function of Biot \cite{biot1956theoryHigh}, whereas we use the nowadays more common Johnson \emph{et al.} model \cite{johnson1987theory}.
Instead of the viscous characteristic length $\Lambda$, the Biot loss function is defined with a characteristic pore size parameter $a$.
By making use of the high-frequency limit of the Biot loss function (equation (3.23) in \cite{biot1956theoryHigh}) we can relate $\Lambda$ to $a$ approximately as

\begin{equation} \label{eq:atoLambda}
\Lambda \approx \dfrac{8k_0\alpha_\infty}{a\phi}.
\end{equation}

\noindent In this paper, we will regard $\Lambda$ from \eqref{eq:atoLambda} as the true value of the viscous length for examining the results, although like mentioned, this is not exact.

To construct a measurement in the frequency domain, we simulate acoustic fields in two configurations for every incidence angle.
The first configuration includes the poroelastic plate, but in the second one the plate is removed and substituted with water to record a reference field.
The transmission coefficient is then obtained as

\begin{equation} \label{eq:transmission}
T^{\mathrm{exp}}(\omega) = \dfrac{p_{T}(\omega)}{p_{\mathrm{ref},T}(\omega)},
\end{equation}

\noindent where $p_{T}$ is the complex pressure recorded at a point on the transmission side of the plate (dot 3 in Fig.~\ref{fig:geometry_2D}), and $p_{\mathrm{ref},T}(\omega)$ is the pressure in the same location without the plate.
To account for the phase difference introduced when the plate is removed and substituted with water, we should multiply equation \eqref{eq:transmission} with a correction factor $\exp\{\I k^{[0]}L\cos\varphi_{\mathrm{inc}}\}$ \cite{jocker2007minimization}.
However, the factor includes parameters $L$ and $\varphi_{\mathrm{inc}}$ that we want to estimate and that are unknown prior to carrying out the inversion.
Therefore during each MCMC iteration, we instead multiply the modelled transmission coefficient \eqref{eq:RTsystem} by an inverse correction factor $\exp\{-\I k^{[0]}L\cos\varphi_{\mathrm{inc}}\}$ and use for $L$ and $\varphi_{\mathrm{inc}}$ the current value of the MCMC chain.

In the configuration with the plate, a microphone on the reflection side records both the incoming and the reflected waves.
Therefore, to calculate the reflection coefficient we need to record two reference locations:

\begin{equation}
R^{\mathrm{exp}}(\omega) = \dfrac{p_{R}(\omega) - p_{\mathrm{inc}}(\omega)}{p_{\mathrm{ref},R}(\omega)},
\end{equation}

\noindent where $p_{R}(\omega)$ is the pressure recorded at a point on the reflection side (dot 1 in Fig.~\ref{fig:geometry_2D}), $p_{\mathrm{inc}}(\omega)$ is the pressure in the same location in the reference configuration, and $p_{\mathrm{ref},R}(\omega)$ is the pressure recorded at a point mirrored on the other side of the reflection surface (dot 2 in Fig.~\ref{fig:geometry_2D}).

For each incidence angle $\varphi_{\mathrm{inc}}$, the simulated reflection and transmission coefficients are solved over $n_\omega$ frequencies, and combined to form a single full measurement $\mathbf{R}^{\mathrm{exp}}(\varphi_{\mathrm{inc}}) = [R^{\mathrm{exp}}(\omega_1;\varphi_{\mathrm{inc}}),\dots,R^{\mathrm{exp}}(\omega_{n_\omega};\varphi_{\mathrm{inc}})]$, and $\mathbf{T}^{\mathrm{exp}}(\varphi_{\mathrm{inc}}) = [T^{\mathrm{exp}}(\omega_1;\varphi_{\mathrm{inc}}),\dots,T^{\mathrm{exp}}(\omega_{n_\omega};\varphi_{\mathrm{inc}})]$.
The measurement vector then reads

\begin{equation}
\mathbf{y}(\varphi_{\mathrm{inc}}) = [\mathbf{R}^{\mathrm{exp}}(\varphi_{\mathrm{inc}}), \mathbf{T}^{\mathrm{exp}}(\varphi_{\mathrm{inc}})]^{T}.
\end{equation}

\noindent Since we mostly deal with inversion from one angle at a time, the $\varphi_{\mathrm{inc}}$-dependence is dropped to improve readability.

\begin{table*}[t]
	\renewcommand{\arraystretch}{1.3}
	\footnotesize
	\caption{Parameters of the prior density}
	\label{tab:prior}
	\begin{tabular}{c|cccccccccccc}
		\hline
		\bfseries Parameter & $\Lambda$ & $\alpha_\infty$ & $\log_{10}k_0$ & $\phi$ & Re $K_b$, $N$ & Im $K_b$, $N$ & Re $K_s$ & Im $K_s$ & $\rho_s$ & $L$ & $\varphi_{\mathrm{inc}}$ & $\sigma_{e_R}$, $\sigma_{e_T}$ \\ \hline
		unit & \multicolumn{1}{c|}{$\mu$m} &  \multicolumn{1}{c|}{-} &  \multicolumn{1}{c|}{m$^2$} &  \multicolumn{1}{c|}{-} & \multicolumn{4}{c|}{GPa} & \multicolumn{1}{c|}{kg$\cdot$m$^{-3}$} & \multicolumn{1}{c|}{mm} & \multicolumn{1}{c|}{deg} & - \\ \hline\hline
		\bfseries Mean & 30 &  2 & $-12$ & 0.4 & 15 & 0 & 30 & 0 & 2500 & 16 & $\varphi_{\mathrm{inc}}^{\mathrm{True}}$ & 0.05 \\
		\bfseries Std & 20 &  1 & 2 & 0.1 & 10 & 0.5 & 20 & 1 & 1000 & 0.2 & 2 & 0.1 \\
		\hline
	\end{tabular}
\end{table*}

\subsection{The noise model}
\label{ssec:noisemodel}

We add Gaussian noise with zero mean to the simulated reflection and transmission coefficients.
Let us denote the noise variance of the reflection and transmission measurements by $\sigma_{e_R}^2$ and $\sigma_{e_T}^2$, respectively.
Level of the additive noise is set to 5 \% of the peak amplitude of the simulated data.
This corresponds to setting the standard deviation $\sigma_{e_R} = 0.05\max\{|\mathbf{R}^{\mathrm{exp}}|\}$.
To achieve the correct noise level for a complex-valued measurement, noise with standard deviation $\sigma_{e_R}/\sqrt{2}$ is added to both the real and imaginary parts of $\mathbf{R}^{\mathrm{exp}}$ (and similarly for $\mathbf{T}^{\mathrm{exp}}$).

The noise level stays constant over the frequency range, which means that the noise covariance can be written as

\begin{equation}
\mathbf{\Gamma}_e = \begin{bmatrix}
\sigma_{e_R}^2 \mathbf{I} & 0 \\
0 & \sigma_{e_T}^2 \mathbf{I} \\
\end{bmatrix},
\end{equation}

\noindent where $\mathbf{I}$ denotes the $n_\omega\times n_\omega$ identity matrix.
The noise level may be estimated from repeated measurements, but sometimes conducting many measurements is not possible or the measurement accuracy is not good enough.
In such a case we often have to make a conservative guess about the noise level, because underestimating the noise variance might mean fitting the model to noise and so the spread of the estimates could be misleading.
The downside is that an overestimated noise level means we are not using all the information the measurements contain.
Therefore we include the noise variances as unknown parameters and estimate them at the same time as the other unknowns.

Now the noise covariance matrix consists of two variables, $\sigma_{e_R}^2$ and $\sigma_{e_T}^2$, and the logarithm of the likelihood density in \eqref{eq:posterior_withnoise} can be written as

\begin{equation} \label{eq:log_likelihood}
\log\pi(\mathbf{y}|\bm{\theta},\sigma_{e_R},\sigma_{e_T}) \propto -\left\Vert \mathbf{L}_e(\mathbf{y} - h(\bm{\theta}))\right\Vert^2 -2n_\omega(\log\sigma_{e_R} + \log\sigma_{e_T}).
\end{equation}

\subsection{The parameter vector and prior}
\label{ssec:prior}

We assume the properties of the saturating fluid to be known, and let the real and imaginary parts of the elastic moduli be independent of each other.
As was mentioned in Sec.~\ref{ssec:Biot_model}, using a model where the real and imaginary parts of the elastic moduli are constant with frequency and inverted independently of each other, leads to slightly non-causal solutions.
Keeping the ratio of imaginary to real part small limits the error coming from the non-causal model.

Including the material thickness, the model now has 12 unknown parameters.
In addition, a measurement of the angle of the incident wave field has some uncertainty as well, so we estimate it with the material parameters.
Hence we have $\bm{\theta} = [\Lambda,\alpha_\infty,k_0,\phi,\mathrm{Re}\:K_b,\mathrm{Im}\:K_b,\mathrm{Re}\:K_s,\mathrm{Im}\:K_s,\mathrm{Re}\:N,\mathrm{Im}\:N,\rho_s,L,\varphi_{\mathrm{inc}}]$.
Adding the noise level parameters, we have 15 unknowns to estimate in total.
Let us denote the parameter vector augmented with the noise level parameters as $\tilde{\bm{\theta}} = [\bm{\theta},\sigma_{e_R},\sigma_{e_T}]$.

We construct the prior using a multivariate normal distribution, i.e. $\tilde{\bm{\theta}}\sim\mathcal{N}(\tilde{\bm{\theta}}_*,\mathbf{\Gamma}_{\tilde{\theta}})$, where $\tilde{\bm{\theta}}_*$ is the prior mean and $\mathbf{\Gamma}_{\tilde{\theta}}$ the covariance.
No correlations between the parameters are assumed, so the prior covariance $\mathbf{\Gamma}_{\tilde{\theta}}$ is diagonal.
To enforce physical limits of the parameters (such as non-negativity), we simply multiply the prior with an indicator function

\begin{equation}
B(\tilde{\theta}_k) =
\begin{cases}
1, & \text{if } \tilde{\theta}_{\mathrm{min},k} \le \tilde{\theta}_k \le \tilde{\theta}_{\mathrm{max},k} \\
0, & \text{otherwise},
\end{cases}
\end{equation}

\noindent where $\tilde{\bm{\theta}}_{\mathrm{min}}$ stands for the minimum, and $\tilde{\bm{\theta}}_{\mathrm{max}}$ for the maximum, admissible values.
Let $\mathbf{L}_{\tilde{\theta}}^T\mathbf{L}_{\tilde{\theta}}^{} = \mathbf{\Gamma}_\theta^{-1}$.
The log posterior can now be stated as

\begin{equation} \label{eq:log_posterior}
\begin{split}
\log\pi(\tilde{\bm{\theta}}|\mathbf{y}) \propto &-\left\Vert \mathbf{L}_e(\mathbf{y} - h(\bm{\theta}))\right\Vert^2 - 2n_\omega(\log\sigma_{e_R} + \log\sigma_{e_T}) \\
& - \frac{1}{2}\Vert \mathbf{L}_{\tilde{\theta}}(\tilde{\bm{\theta}} - \tilde{\bm{\theta}}_*)\Vert^2 + \log B(\tilde{\bm{\theta}}).
\end{split}
\end{equation}

Most of the parameters have $\tilde{\theta}_{\mathrm{min},k} = 0$ and $\tilde{\theta}_{\mathrm{max},k} = \infty$, corresponding to a non-negativity constraint.
The exceptions to this are the imaginary parts of the elastic moduli which are non-positively constrained, and porosity, tortuosity, and permeability.
Porosity is, by definition, constrained between $\phi\in[0,1]$, and tortuosity is only defined for $\alpha_\infty \ge 1$.
Permeability on the other hand has possible values that span multiple orders of magnitude.
The range between the least and most permeable porous media \cite{schon2015physical} can be up to $k_0\in [10^{-16},10^{-8}]$.
For the inversion we therefore prefer to transform $k_0$ to the logarithmic scale, $\log_{10}(k_0)\in[-16,-8]$, and only transform it back to the real scale during the forward model evaluation.

In a numerical study the true parameter values are known, and we need to be careful not to exploit this information when constructing the prior density.
We have attempted to select the prior mean and standard deviation conservatively, reflecting the kind of values we could reasonably expect in the given type of materials (see Table~\ref{tab:prior}).
However, we assume that we can measure thickness and the incidence angle reasonably accurately and reflect that by a smaller standard deviation.

In addition, we impose a constraint on $K_b$ based on the Biot-Willis coefficient $\alpha_{\mathrm{B}}$ \cite{biot1957elastic}, which is bounded by $\phi~\le~\alpha_{\mathrm{B}}~\le~1$.
The lower limit corresponds to a rigid frame, and the upper limit to a soft or limp frame.
The upper limit is enforced by the positivity constraint of $K_b$, but for the lower limit we require $K_b \le K_s(1 - \phi)$.
This constraint has little effect on the posterior density, but considerably limits the volume of the prior space, and thus we need fewer temperatures to bridge the gap between the coldest and hottest chains.
We implement this constraint using the indicator function as well.

\begin{figure*}[t]
	\includegraphics[width=\linewidth]{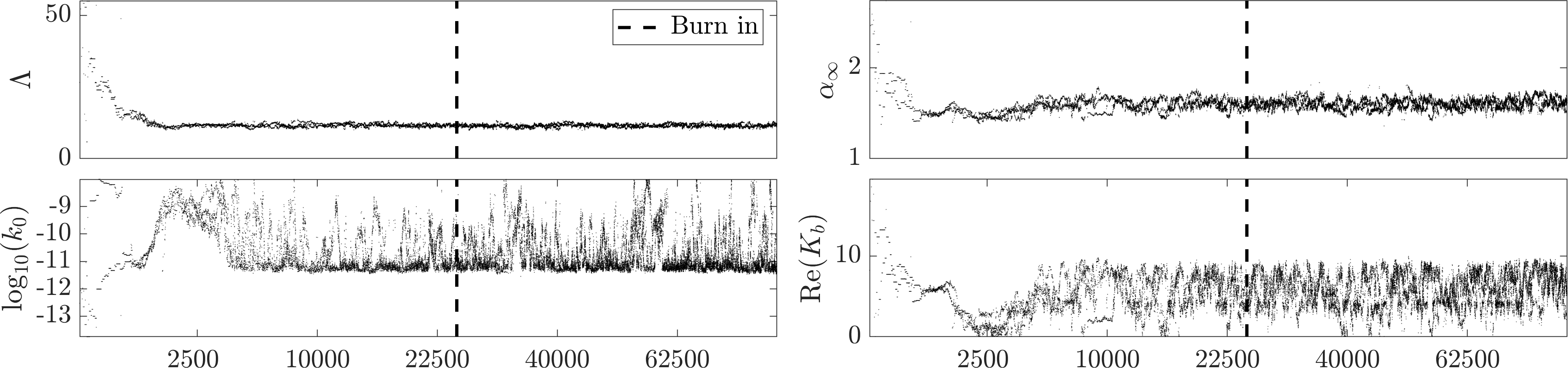}
	\caption{Posterior chains of $\Lambda$, $\alpha_\infty$, $\log_{10}(k_0)$, and $Re(K_b)$. Incidence angle 0 $^{\circ}$. Note the $x$-axis is plotted on a $\sqrt{x}$ scale to better show the beginning of the chain.}
	\label{fig:TS_0deg}
\end{figure*}

\subsection{MCMC sampling}

Before presenting the results let us briefly comment on the sampling process.
We show the case $\varphi_{\mathrm{inc}} = 0 ^{\circ}$ as an example, and just note that the MCMC algorithm works similarly at other angles of incidence unless stated otherwise.
Results for measurements at normal incidence are probably the most interesting since normal incidence corresponds to the easiest (and hence the most common) measurement setup.

A starting location for each chain is drawn from the prior.
The algorithm is insensitive to the starting location, i.e. we are able to find convergence no matter what the initial point is.
The first 25,000 samples are discarded as burn-in, during which we can observe the posterior chain (the chain with $T = 1$) reaching a more or less stable state, see Fig.~\ref{fig:TS_0deg}.
The big jumps that occur in the beginning of the simulation, and bring the chain quickly to the main posterior mode, are typical for the parallel tempering algorithm.
We can also observe how the proposal covariance adapts to the target density during the burn-in.
To achieve a good between-chain swap acceptance rate, we used 11 temperatures.
The stopping criterion was reached at 75,000 iterations.

\begin{figure*}[t]
	\includegraphics[width=\linewidth]{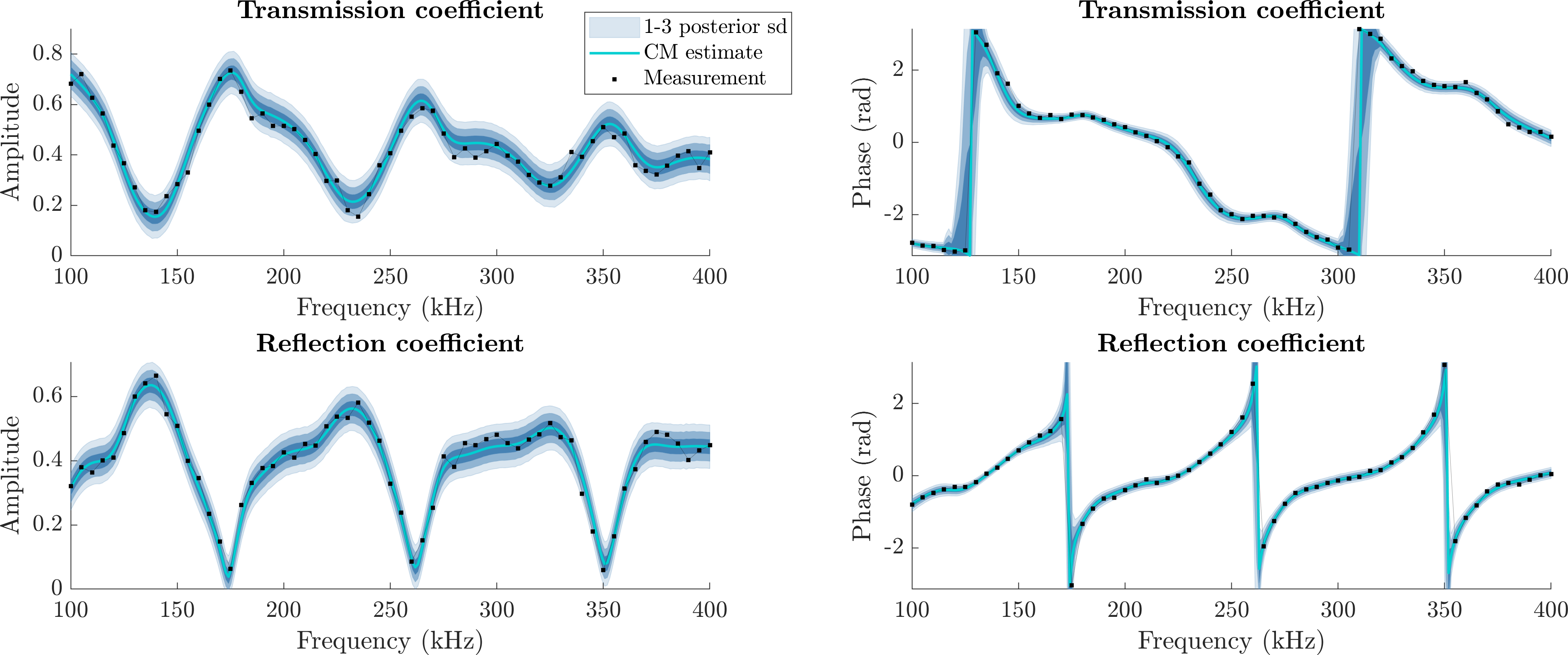}
	\caption{A Comsol measurement with added noise, the model prediction corresponding to the CM estimate, and the posterior predictive distribution as 1-3 standard deviations. Case of $\varphi_{\mathrm{inc}} = 0$ deg.}
	\label{fig:0deg_PE3_CM}
\end{figure*}

As a model feasibility check, we plot the posterior predictive distribution (pdd) in Fig.~\ref{fig:0deg_PE3_CM}.
It shows the noisy measurement data, the CM estimate, and the pdd as standard deviations of the model predictions.
Standard deviations are a good way to summarise the pdd since we found it approximately normally distributed.
The graphical check shows that there are no systematic discrepancies between the Comsol data and predictions from our model at normal incidence.

Lastly we note that sampling the posterior at 20 and 25 degrees of incidence turned out to be more difficult than at other angles.
We needed 75,000 samples and 13 temperatures to reach a stable state, and convergence was achieved at 150,000 samples.
One explanation for this is that the ratio between the width of the posterior and prior density is more extreme at these angles, and thus the temperature steps need to be smaller.

\section{Results}
\label{sec:results}

\begin{figure*}[t]
	\includegraphics[width=\linewidth]{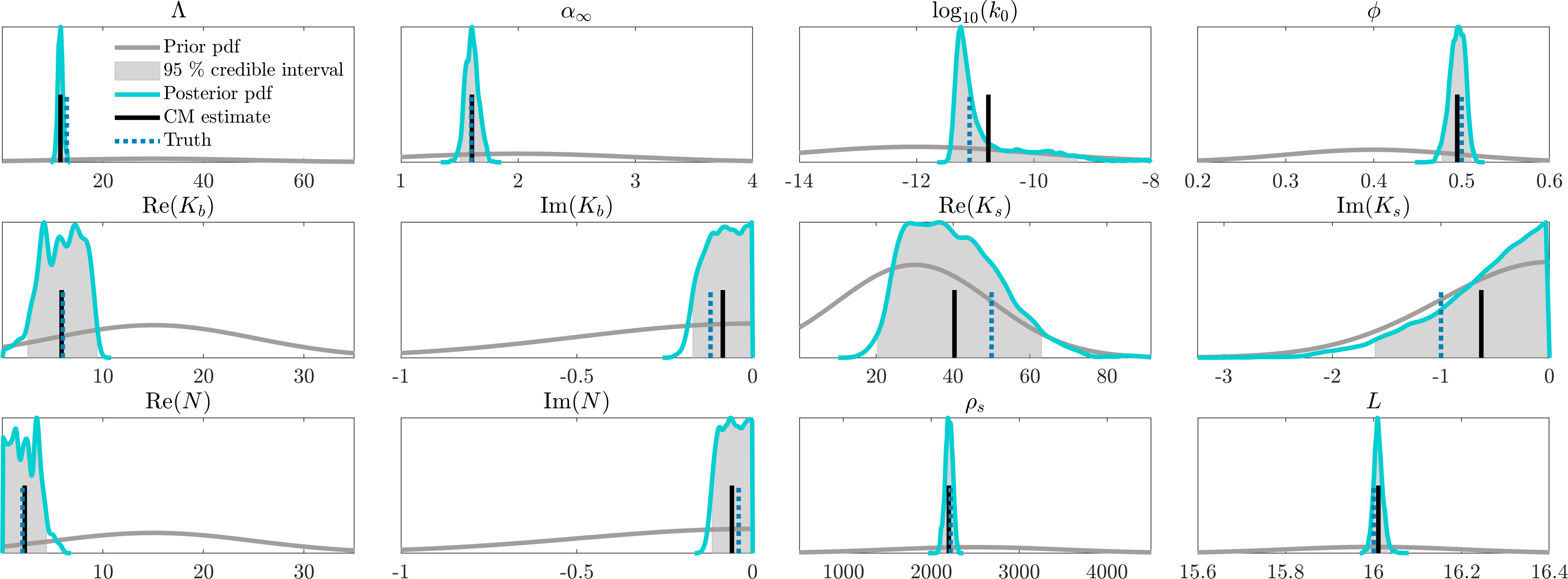}
	\caption{Marginal posterior and prior probability distributions for $\varphi_{\mathrm{inc}} = 0^{\circ}$. The posterior and prior are normalised so they integrate to one. The full posterior and at least 95 \% of the prior mass is shown to enable comparison.}
	\label{fig:0deg_PE3_1D}
\end{figure*}

Let us now examine the inversion results, again starting from normal incidence.
Fig.~\ref{fig:0deg_PE3_1D} shows the marginal distributions of the posterior, along with the prior densities.
The shaded areas denote the 95 \% credible posterior intervals.
Comparing the width of the marginal posterior and prior distributions tells about the amount of information provided by the measurement data (the narrower the posterior, the more informative the measurement).
A visual inspection shows that the posterior densities of $\Lambda, \alpha_\infty, \phi, \rho_s$, and $L$ are the most accurate by this criterion.

The posterior density of permeability has a peak at its true value, but the distribution continues to the specified upper limit of $10^{-8}$ m$^{2}$.
However, there seems to be a clear lower bound below which $k_0$ cannot admit values.
Hence the CM estimate tends to overestimate permeability.
Moreover, this shows why it is advisable to express $k_0$ on a log scale.
On a linear scale the values near the upper bound would completely dominate the distribution and we would not be able to see the peak at the true parameter value.

The elastic parameters are not as well identified as are the other unknowns.
We can even see several peaks in the distribution of $K_b$ and $N$.
The real and imaginary parts of $K_s$ are almost not identified at all and their posterior distribution mainly follows the prior.
The real and imaginary parts of $K_b$ and $N$ are better identified, although they have lots of uncertainty and the 95~\% credible interval even admits values down to zero.

\begin{figure}[t!]
	\includegraphics[width=1\linewidth]{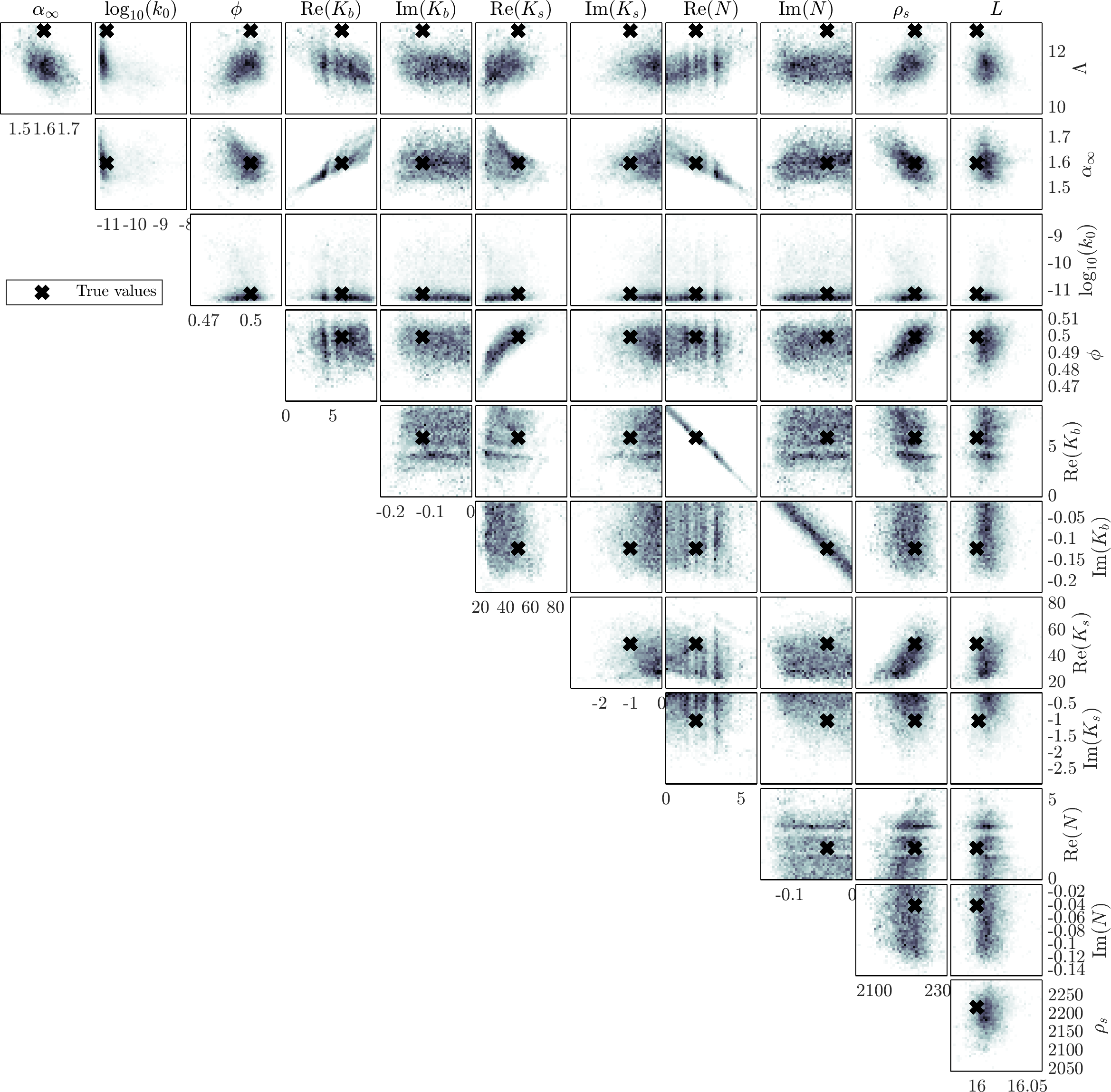}
	\caption{Joint marginal posterior distributions, $\varphi_{\mathrm{inc}} = 0^{\circ}$.}
	\label{fig:0deg_PE3_2D}
\end{figure}

Checking the joint marginal distributions provides complementary information about the problem of identifiability and non-identifiability.
These distributions are shown in Fig.~\ref{fig:0deg_PE3_2D}.
We can immediately point out a strong negative linear correlation between the real (and to an extent the imaginary) parts of $K_b$ and $N$.
The strong negative correlation between $K_b$ and $N$ explains why the parameters are able to have values near zero.
To understand where the correlation comes from, we recall that at normal incidence no shear waves are produced and therefore we do not see their effect.
However, the fast longitudinal, or the P-, wave is controlled by both the bulk and shear modulus as illustrated \cite{jocker2009ultrasonic} by the P-wave speed of a drained porous frame ($K_f \ll K_b,N$):

\begin{equation} \label{eq:drainedPspeed}
V_p = \sqrt{\dfrac{K_b + \frac{4}{3}N}{(1 - \phi)\rho_s}}.
\end{equation}

\noindent While our material is saturated with fluid that has $K_f \approx N$, the difference between \eqref{eq:drainedPspeed} and the saturated material P-wave speed (see for example \cite{johnson1994probing}) is less than 1 \% in the considered frequency range and we can use the simple expression \eqref{eq:drainedPspeed} to give physical intuition.
At the same time the only expression where $N$ can be found without $K_b$ is the expression for the shear wave velocity:

\begin{equation}
V_{sh} = \sqrt{\frac{N}{\rho - \rho_f^2/\tilde{\rho}_{\mathrm{eq}}}}.
\end{equation}

\noindent This implies that we need to have shear waves to find both $K_b$ and $N$, and in normal incidence can only find a linear combination of $K_b$ and $N$.

Tortuosity is another parameter that is highly correlated with the elastic constants.
This is also probably due to a lack of information about the wave speeds, because tortuosity determines the speed of the slow longitudinal wave.
The good news is that the true parameter values are all found on the narrow and long support of the joint densities.
This means that given accurate prior information on one of the highly correlated parameters, we can expect to estimate the other correlated parameters more accurately as well.

\subsection{Oblique angles}

\begin{figure}[t]
	\includegraphics[width=\linewidth]{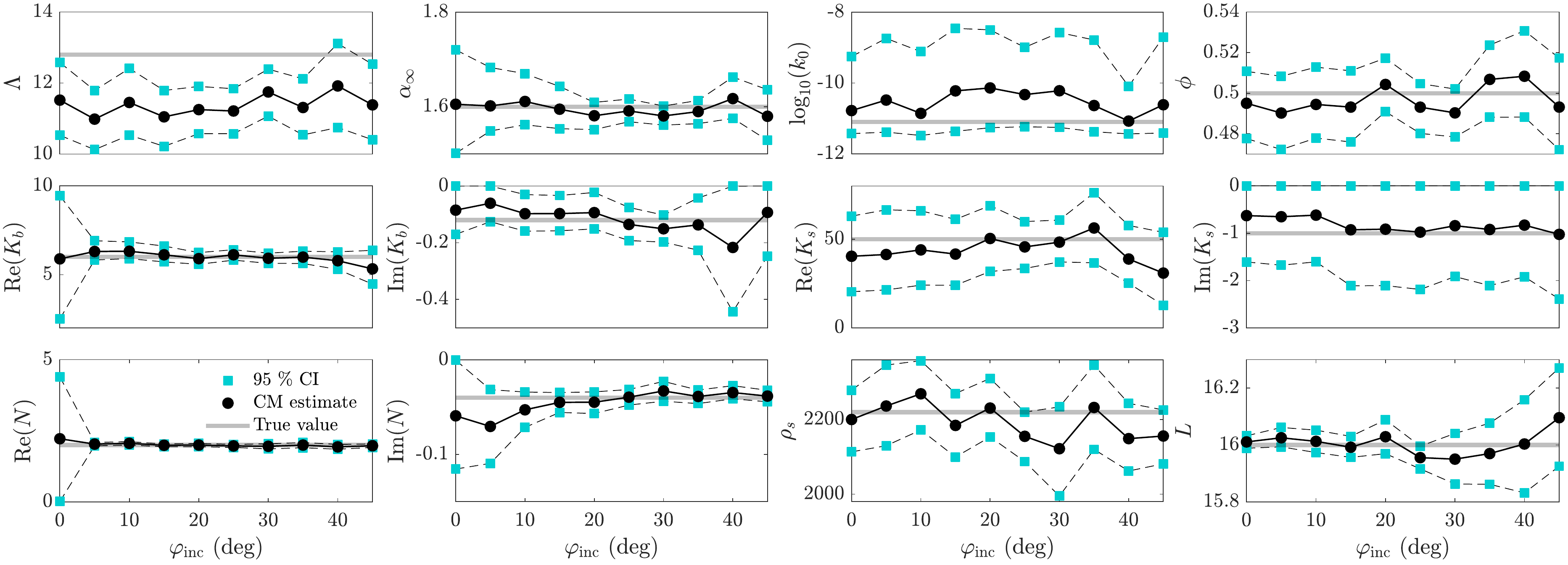}
	\caption{The estimated parameters and 95 \% credible intervals as a function of the measurement angle.}
	\label{fig:params_angles}
\end{figure}

Fig.~\ref{fig:params_angles} shows the CM estimates and 95 \% credible intervals of each parameter as a function of the incidence angle from 0 to 45 degrees.
We can see that the credible intervals contain the true parameter values at each angle, with the exception of characteristic viscous length.
The marginal posterior densities for the two extreme angles, 0 and 45 degrees, are summarised numerically in Table~\ref{tab:inversion_results}.

The results for oblique angles are as expected in that the estimation of the bulk and shear modulus of the frame is remarkably more accurate than at normal incidence.
Now the measurements include effects of the shear wave, and it is thus possible to estimate $N$ and $K_b$ separately.
In addition, tortuosity is estimated more accurately at higher angles of incidence.
Accuracy of the other parameters stays approximately unchanged, except that, interestingly, the imaginary part of $N$ gets more accurate but the imaginary part of $K_b$ gets less accurate as $\varphi_{\mathrm{inc}}$ increases.

Three parameters, porosity, viscous length, and bulk modulus of the solid, seem to be underestimated slightly at most of the angles.
For porosity and solid bulk modulus the prior mean is set lower than the true value, and it might be pulling the posterior below the true value.
They are also positively correlated, so that a lower porosity can be compensated by a lower bulk modulus.
However, the fact that $\Lambda$ is consistently underestimated shows that the expression \eqref{eq:atoLambda} used to find the ''true'' value is not exact.
We confirmed this by considering a noiseless measurement, and maximising the posterior \eqref{eq:posterior} with a Gauss-Newton iteration that was started at the true values.
We found that while other parameters stayed at their starting location, the best fit was achieved with $\Lambda = 11.6$ $\mu$m, a value that also corresponds to the CM estimates.

The only parameter that clearly gets less accurate as the angle increases is the thickness of the plate.
This also makes sense because the amplitudes of reflections inside the plate, which inform about the thickness, are smaller at higher angles and thus more easily masked by noise.

\begin{table}[h]
	\renewcommand{\arraystretch}{1.3}
	\footnotesize
	\caption{Conditional mean (CM) and 95 \% credible interval estimates.}
	\label{tab:inversion_results}
	\centering
	\begin{tabular}{c||c|l@{ }c@{ }r|l@{ }c@{ }r}
		\multicolumn{2}{c|}{} & \multicolumn{3}{c|}{$\varphi_{\mathrm{inc}} = 0\:^{\circ}$} & \multicolumn{3}{c}{$\varphi_{\mathrm{inc}} = 45\:^{\circ}$} \\
		
		\hline
		\bfseries Parameter & \bfseries Truth & \multicolumn{3}{c|}{\bfseries [$a_I$, CM, $b_I$]$_{95 \%}$} & \multicolumn{3}{c}{\bfseries [$a_I$, CM, $b_I$]$_{95 \%}$} \\
		
		\hline\hline
		$\Lambda$ & 12.8 * & [10.5,&\textbf{11.5},&12.6] & [10.4,&\textbf{11.4},&12.5] \\
		
		$\alpha_\infty$ & 1.6 & [1.50,&\textbf{1.61},&1.72] & [1.53,&\textbf{1.58},&1.64] \\
		
		$\log_{10}k_0$ & -11.1 & [-11.4,&\textbf{-10.8},&-9.3] & [-11.4,&\textbf{-10.6},&-8.7] \\
		
		$\phi$ & 0.5 & [0.48,&\textbf{0.50},&0.51] & [0.47,&\textbf{0.49},&0.52] \\
		
		Re $K_b$ & 6 & [2.51,&\textbf{5.89},&9.44] & [4.47,&\textbf{5.33},&6.37] \\
		
		Im $K_b$ & -0.12 & [-0.17,&\textbf{-0.09},&-0.00] & [-0.25,&\textbf{-0.09},&-0.00] \\
		
		Re $K_s$ & 50 & [20.3,&\textbf{40.4},&63.0] & [12.7,&\textbf{30.9},&53.9] \\
		
		Im $K_s$ & -1 & [-1.61,&\textbf{-0.63},&-0.00] & [-2.39,&\textbf{-1.02},&-0.00] \\
		
		Re $N$ & 2 & [0.01,&\textbf{2.22},&4.40] & [1.91,&\textbf{1.97},&2.03] \\
		
		Im $N$ & -0.04 & [-0.12,&\textbf{-0.06},&-0.00] & [-0.04,&\textbf{-0.04},&-0.03] \\
		
		$\rho_s$ & 2220 & [2113,&\textbf{2200},&2279] & [2081,&\textbf{2156},&2226] \\
		
		$L$ & 16 & [15.99,&\textbf{16.01},&16.04] & [15.93,&\textbf{16.10},&16.27] \\
		
		\hline
	\end{tabular} \\
	\vspace{0.5mm}\flushleft
	*Calculated from \eqref{eq:atoLambda} with $a = 16 \:\mu$m.
\end{table}

\begin{figure}[t!]
	\includegraphics[width=\linewidth]{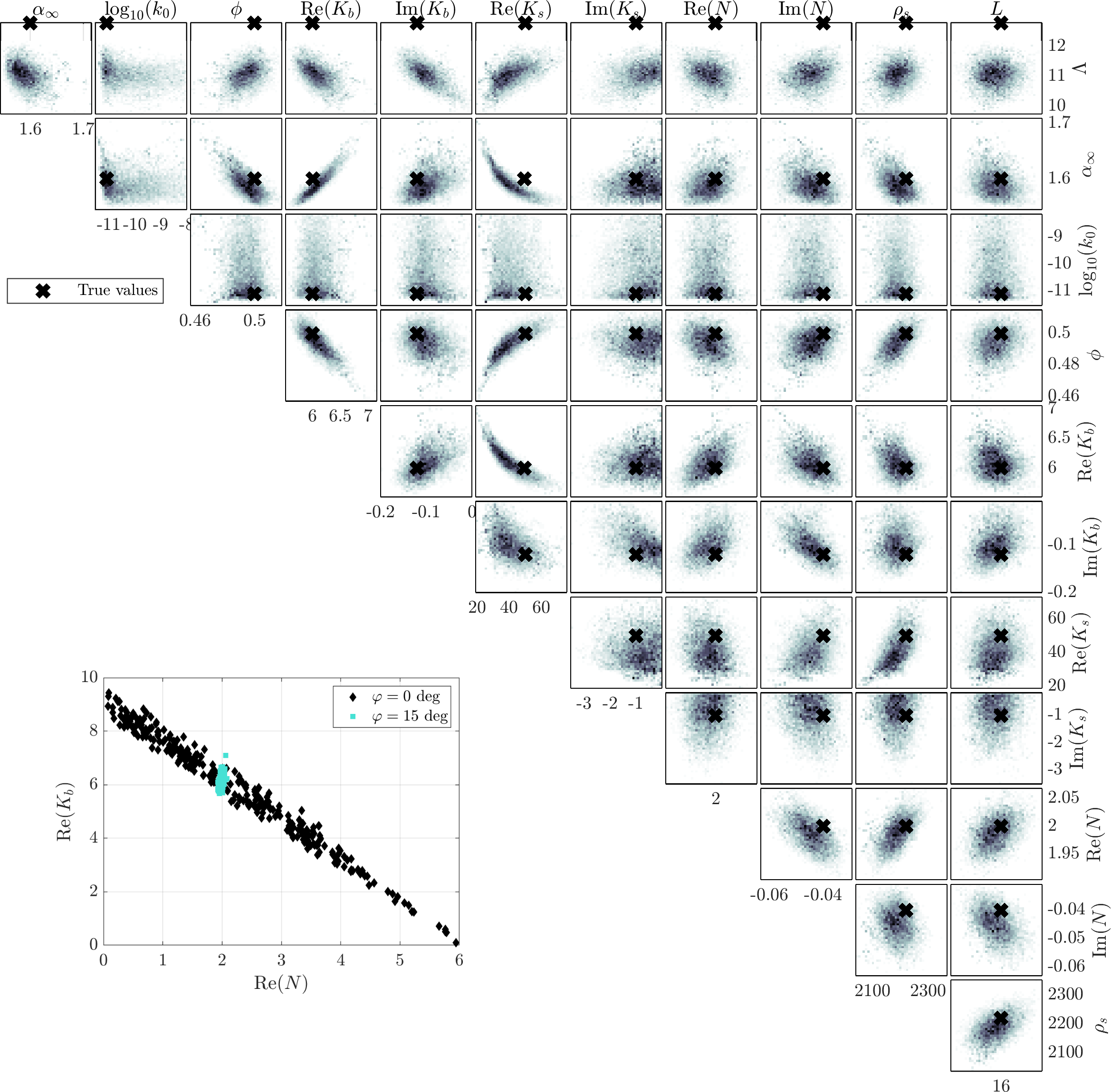}
	\caption{Joint marginal posterior distributions for $\varphi_{\mathrm{inc}} = 15\:^{\circ}$, and a comparison to the normal incidence case.}
	\label{fig:15deg_PE3_2D}
\end{figure}

Let us take another look at the joint marginal distributions (Fig.~\ref{fig:15deg_PE3_2D}), but with an oblique angle of 15 degrees this time.
The first observation is that $K_b$ and $N$ are no longer nearly as correlated, and specifically, the shear modulus does not exhibit any of the previous correlations.
However, since some of the parameters are more accurately resolved, we start to see correlations that were not visible in the normal incidence case.
For example $N$ and $\rho_s$ show moderate correlation, which is reasonable since they both control the speed of the shear wave.
In addition, strong correlations can be seen between $K_s$ and tortuosity, porosity, and $K_b$.
This shows that the solid bulk modulus is not very well identified even from an oblique incidence data, and that its effect can be compensated by tuning several other parameters.
To achieve an accurate result, we would need to either measure $K_s$ separately and use it as prior information, or devise a new kind of measurement set up altogether.
The comparison in the lower left corner of Fig.~\ref{fig:15deg_PE3_2D} shows how much smaller the joint posterior is for $N$ and $K_b$ in the 15 degree situation.

\begin{table}[h]
	\renewcommand{\arraystretch}{1.3}
	\footnotesize
	\caption{Conditional mean (CM) and 95 \% credible interval estimates.}
	\label{tab:inversion_results_v2}
	\centering
	\begin{tabular}{c||c|l@{ }c@{ }r|l@{ }c@{ }r}
		\multicolumn{2}{c|}{} & \multicolumn{3}{c|}{$\varphi_{\mathrm{inc}} = 0$ and $15\:^{\circ}$} & \multicolumn{3}{c}{$\varphi_{\mathrm{inc}} = 10\:^{\circ}\rightarrow 8\:^{\circ}$} \\
		
		\hline
		\bfseries Parameter & \bfseries Truth & \multicolumn{3}{c|}{\bfseries [$a_I$, CM, $b_I$]$_{95 \%}$} & \multicolumn{3}{c}{\bfseries [$a_I$, CM, $b_I$]$_{95 \%}$} \\
		
		\hline\hline
		$\Lambda$ & 12.8 * & [10.3,&\textbf{11.0},&11.5] & [10.1,&\textbf{11.2},&12.5] \\
		
		$\alpha_\infty$ & 1.6 & [1.56,&\textbf{1.59},&1.62] & [1.57,&\textbf{1.66},&1.75] \\
		
		$\log_{10}k_0$ & -11.1 & [-11.2,&\textbf{-10.1},&-8.4] & [-11.4,&\textbf{-10.3},&-8.6] \\
		
		$\phi$ & 0.5 & [0.48,&\textbf{0.50},&0.51] & [0.45,&\textbf{0.47},&0.49] \\
		
		Re $K_b$ & 6 & [5.81,&\textbf{6.16},&6.52] & [5.92,&\textbf{6.62},&7.28] \\
		
		Im $K_b$ & -0.12 & [-0.14,&\textbf{-0.10},&-0.05] & [-0.25,&\textbf{-0.17},&-0.08] \\
		
		Re $K_s$ & 50 & [27.8,&\textbf{46.1},&68.0] & [17.4,&\textbf{30.8},&52.5] \\
		
		Im $K_s$ & -1 & [-1.98,&\textbf{-0.93},&-0.00] & [-2.04,&\textbf{-0.86},&-0.00] \\
		
		Re $N$ & 2 & [1.98,&\textbf{2.01},&2.04] & [1.91,&\textbf{1.96},&2.00] \\
		
		Im $N$ & -0.04 & [-0.05,&\textbf{-0.05},&-0.04] & [-0.03,&\textbf{-0.02},&-0.01] \\
		
		$\rho_s$ & 2220 & [2172&\textbf{2229},&2282] & [1995,&\textbf{2071},&2151] \\
		
		$L$ & 16 & [15.99,&\textbf{16.00},&16.02] & [15.91,&\textbf{15.93},&15.95] \\
		
		\hline
	\end{tabular} \\
	\vspace{0.5mm}\flushleft
	*Calculated from \eqref{eq:atoLambda} with $a = 16 \:\mu$m.
\end{table}

\section{Discussion}
\label{sec:discussion}

To extract the maximum amount of information possible from acoustic reflection and transmission measurements, it is recommended to measure the poroelastic object at an oblique angle.
This way the shear wave effects are recorded.
At the same time, measurements at normal incidence are the simplest to perform and are enough to estimate many parameters, such as porosity, permeability, density of the solid part, and viscous length, with accuracy equal to the oblique angle measurements.
Normal incidence may be all that is needed for example in reservoir characterisation, where the two main microscopic-scale properties of interest are porosity and permeability \cite{slatt2013stratigraphic}.

We can of course join several measurements together for the inversion.
This way complementary information from different angles can be used at the same time.
Results from inversion using normal incidence and 15 degrees together as data are shown in Table~\ref{tab:inversion_results_v2}.
We can see that the results are very accurate.

In the case we consider in this paper, the minimum angle required to include the shear effects into inversion is only five degrees.
However, this is not a universal truth, and for several reasons will likely be case dependent.
In real world experiments, we may have to deal with a larger model error, i.e. there will be bigger differences between the (ideal, noiseless) measurement and the inverse solution.
In this work, sources of difference between our model and the Comsol implementation include distortions from the truncation of the computational domain, and a different viscosity model.
However, the solutions returned by the forward model agree well with the finite element model, and we conclude that the modelling errors in this case are relatively small.

On the other hand, the level of additive noise used in this work, five percent, is quite large compared to what can be achieved by repeated measurements in laboratory settings.
The advantage of having a noise dominated measurement as opposed to model error dominated one in the inversion is that we know (or at least can reasonably assume) the structure of the measurement noise.
The noise level naturally affects the minimum angle needed to resolve shear effects.
For example, from a simulation with ten percent additive noise, we found that in order to estimate both $K_b$ and $N$ accurately, the incident wave angle has to be at least ten degrees.
We also achieved similar results as in the ten percent noise case by keeping the noise level at five percent but halving the amount of measurement points to 31 frequencies.
Another observation regarding the noise level is that the more there is noise, the easier the inverse problem is to solve from a sampling point of view.
With small noise levels the likelihood is narrower and more peaky, which is obviously harder to sample and requires more temperatures.

In this ultrasonic measurement setup, we can estimate, for example, porosity to within five percent of its true value, and the true value happens to be always included within the 95 percent credible intervals.
Permeability is more difficult to find accurately, even with the transformation to a logarithmic scale.
For the considered material we do find a lower bound, which also corresponds closely to the true value.
However, the likelihood does not seem to provide an upper bound for permeability.
The undefined upper bound is a symptom of the fact that in the current model, reflection and transmission coefficients do not change when permeability is higher than its true value.

In summary, the preferred angle of incidence is dependent on the parameters in which we are the most interested.
Most of the parameters, and a linear combination of $K_b$ and $N$ can be estimated reliably from just the normal incidence measurement, but using a measurement from an oblique angle allows us to separate $N$ from $K_b$.
Another option is to measure the shear modulus separately, in which case normal incidence would be enough to invert everything else, but this requires another experimental system.

\subsection{Uncertainty of measurement conditions}

Estimation accuracy of the material parameters depends on several factors, such as noise level, accuracy of prior knowledge, validity of models, and sampling accuracy.
An often overlooked source of error is the uncertainty in the measurement setup.
This includes variables such as the angle of the incident wave, any dimensions of the measurement system and the measured object that we use, and possibly even the level of measurement noise.
We have not found attempts to control for these in the literature concerning characterisation of porous media.

Ignoring uncertainty in the angle of incidence produces misleading estimates.
The incidence angle exhibits correlation with several parameters, such as material thickness, density, and some of the elastic parameters.
This means that if the angle uncertainty is not accounted for, distributions of the parameters that are correlated with the angle will be too narrow and, as a result, the credible intervals may not include the true parameter values.
To see the effect of a slightly wrong fixed value for the incidence angle, we ran a test at $\varphi_{\mathrm{inc}} = 10$ degrees, and set the incidence angle slightly wrong in the inversion, to 8 degrees.
The results are shown in Table~\ref{tab:inversion_results_v2}.
We notice that $\rho_s$, $L$, and $\phi$ are already underestimated by so much that the true value is not within the 95 percent CI.

\begin{figure}[t!]
	\includegraphics[width=\linewidth]{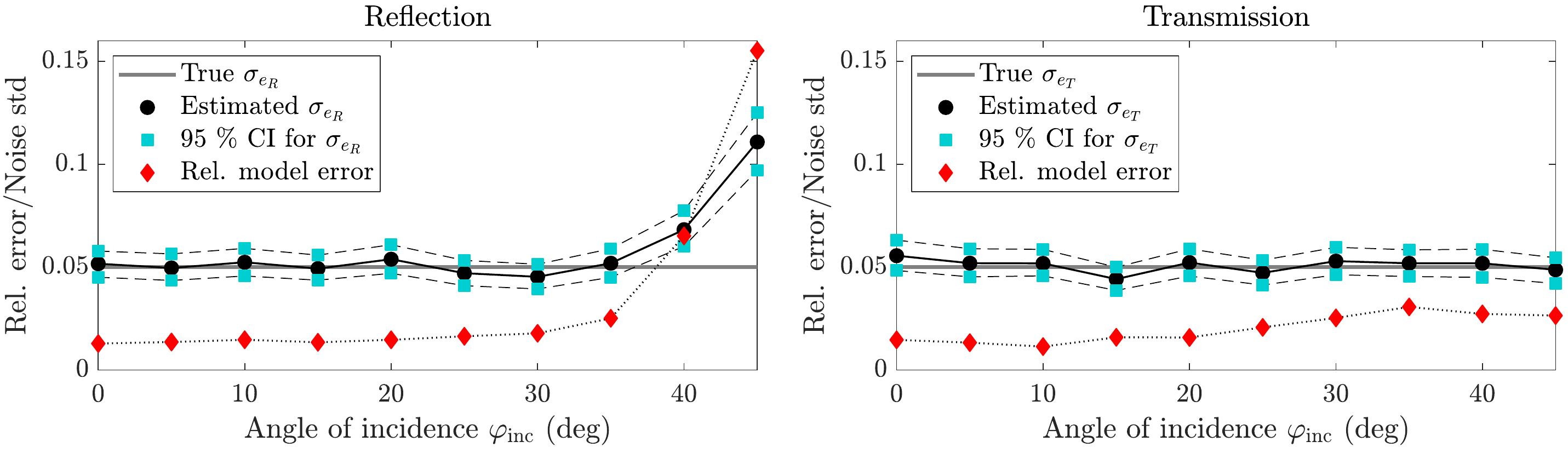}
	\caption{Estimated noise standard deviations as a function of incidence angle. Also included is the relative error between the outputs of Comsol and GMM models.}
	\label{fig:errors_angles}
\end{figure}

Another parameter not directly related to the material properties is variance of the noise, in both the reflection and transmission measurements.
Inversion results for the noise parameters are shown in Fig.~\ref{fig:errors_angles}.
We find that $\sigma_{e_T}$ is correctly estimated for all considered angles of incidence, but the estimate of $\sigma_{e_R}$ seems to be too big for angles past 30 degrees.
However, we also notice that at higher angles the reflection part of the GMM model does not fit the COMSOL measurements as well as it does in the lower angles.
This can be seen by plotting the relative model error, defined by (resp. for $\mathbf{T}$)

\begin{equation}
\mathrm{Rel.\:model\:error} = \frac{\left\Vert \mathbf{R}^{\mathrm{exp}} - \mathbf{R}^{\mathrm{GMM}}\right\Vert}{\left\Vert \mathbf{R}^{\mathrm{exp}}\right\Vert},
\end{equation}

\noindent where the coefficients are calculated with the same parameter values and no noise is added.
The model error that increases with the incidence angle is caused by our choice of an imperfect absorbing boundary condition in the finite element model.
We use an impedance boundary, which is known to work the best only when the boundary is normal to an incoming wave.

We can see in Fig.~\ref{fig:errors_angles} that the overestimated $\sigma_{e_R}$ seems to be compensating for the model error.
In this way, the model error is approximated with the same distribution that the measurement noise has, which is not necessarily a good approximation.
Nevertheless, when we do not estimate the noise level and instead fix it at the correct value of five percent, we find that the estimates at angles where there is model error are wrong.
For example, the estimate for permeability was found to be much narrower than in other cases and the 95 percent credible interval did not even contain the true value.

\section{Conclusion}
\label{sec:conclusion}

In this paper we examined the feasibility of the inverse characterisation of poroelastic media based on simulated ultrasonic measurement data.
We considered a situation where a slab of poroelastic material is submerged in water and measured from different angles.
First we solved the alternative formulation of the Biot equations for poroelastic media to find the acoustic reflection and transmission coefficients of the object of interest.
To increase stability of the solution, we used the global matrix method instead of the usual Thomson-Haskell transfer matrix method.

The inverse problem was formulated in the Bayesian framework to take advantage of prior information, account for measurement uncertainties, and to be able to straightforwardly assess the uncertainty in the parameter estimates.
We used an adaptive parallel tempering MCMC algorithm to sample the solution of the inverse problem, the posterior probability density.
Such a sampling approach was necessary because the posterior density was found to be highly peaked, and standard single-chain samplers tended to get stuck and not explore the posterior in a satisfactory way.

We showed that the proposed approach is able to extract reliably the information contained by the acoustic transmission and reflection measurements.
Measurements at normal incidence do not provide information on the shear properties of the material, but small oblique incidences can be enough to get that information.
It is preferable to account for uncertainty in the angle of the incident wave to correctly estimate the parameters.
In addition, we found that it is possible to estimate accurately the level of measurement noise, and small modelling errors can also be compensated for using this approach.

\section*{Acknowledgment}
This work has been supported by the strategic funding of the University of Eastern Finland, by the Academy of Finland (Finnish Centre of Excellence of Inverse Modelling and Imaging), and by the RFI Le Mans Acoustique (Pays de la Loire) Decimap project. This article is based upon work initiated under the support from COST Action DENORMS CA-15125, funded by COST (European Cooperation in Science and Technology).

\appendix
\section{The state matrix} \label{app:statematrix}

The state matrix is of the form $\mathbf{A} = \mathbf{B}^{-1}\mathbf{D}$, with (after Gautier \emph{et al.} \cite{gautier2011propagation})

\begin{equation}
\mathbf{B} = \begin{bmatrix}
1 & 0 & 0 & 0 & \I k_1(\lambda_c - \alpha_{\mathrm{B}} M) & -\dfrac{k_1\alpha_{\mathrm{B}} M}{\omega} \\
0 & 1 & 0 & 0 & 0 & 0 \\
0 & 0 & -1 & 0 & 0 & 0 \\
0 & 0 & 0 & N & 0 & 0 \\
0 & 0 & 0 & 0 & \lambda_c + 2N - \alpha_{\mathrm{B}} M & \I\dfrac{\alpha_{\mathrm{B}} M}{\omega} \\
0 & 0 & 0 & 0 & (\alpha_{\mathrm{B}} - 1)M & \I\dfrac{M}{\omega} \\
\end{bmatrix},
\end{equation}

\noindent and

\begin{equation}
\mathbf{D} = \begin{bmatrix}
0 & 0 & \I k_1\left(\dfrac{\rho_f}{\tilde{\rho}_{\mathrm{eq}}} - \dfrac{k_1^2\alpha_{\mathrm{B}} M}{\tilde{\rho}_{\mathrm{eq}}\omega^2}\right) & \omega^2\left(\rho - \dfrac{\rho_f^2}{\tilde{\rho}_{\mathrm{eq}}}\right) - k_1^2\left(\lambda_c + 2N - \dfrac{\rho_f}{\tilde{\rho}_{\mathrm{eq}}}\alpha_{\mathrm{B}} M\right) & 0 & 0 \\
\I k_1 & 0 & 0 & 0 & (\rho - \rho_f)\omega^2 & \I\rho_f\omega \\
0 & 0 & 0 & 0 & (\rho_f - \tilde{\rho}_{\mathrm{eq}})\omega^2 & \I\tilde{\rho}_{\mathrm{eq}}\omega \\
-1 & 0 & 0 & 0 & \I k_1N & 0 \\
0 & -1 & -\dfrac{k_1^2\alpha_{\mathrm{B}} M}{\tilde{\rho}_{\mathrm{eq}}\omega^2} & \I k_1\left(\lambda_c - \dfrac{\rho_f}{\tilde{\rho}_{\mathrm{eq}}}\alpha_{\mathrm{B}} M\right) & 0 & 0 \\
0 & 0 & 1 - \dfrac{k_1^2M}{\tilde{\rho}_{\mathrm{eq}}\omega^2} & \I k_1M\left(\alpha_{\mathrm{B}} - \dfrac{\rho_f}{\tilde{\rho}_{\mathrm{eq}}}\right) & 0 & 0 \\
\end{bmatrix}.
\end{equation}

\bibliographystyle{unsrt}

\end{document}